\documentclass[extra, mreferee]{gji}
\usepackage{epsfig, color}
\usepackage{amssymb, amsmath}
\usepackage{setspace}

\title[Waves in space-dependent and time-dependent materials]
 {Waves in space-dependent and time-dependent materials: a systematic comparison}

\author[Wapenaar et al.]
{\small Kees Wapenaar$^1$, Johannes Aichele$^2$ and Dirk-Jan van Manen$^2$\\
$^1$Department of Geoscience and Engineering, Delft University of Technology, The Netherlands\\
$^2$ETH Z\"urich, Switzerland}
\pagerange{1--99}
\volume{999}
\pubyear{2023}

\begin{document}
\begin{spacing}{2.0}
\label{firstpage}

\maketitle

\begin{summary}

{\small

Waves in space-dependent and in time-dependent materials obey similar wave equations, with interchanged time- and space-coordinates.
However, since the causality conditions are the same in both types of material (i.e., without interchangement of time- and space-coordinates), 
the solutions are dissimilar.

We present a systematic treatment of wave propagation and scattering in 1D space-dependent and in 1D time-dependent materials.
After formulating unified equations, we discuss Green's functions 
and simple wave field  representations for both types of material.
Next we discuss propagation invariants, i.e., quantities that are independent of the space coordinate 
in a space-dependent material (such as the net power-flux density) or of the time coordinate in a time-dependent material (such as the net field-momentum density).
A discussion of general reciprocity theorems leads to the well-known source-receiver reciprocity relation for the Green's function of a space-dependent material and a new
source-receiver reciprocity relation for the Green's function of a time-dependent material.
A discussion of general wave field representations leads to the well-known expression for Green's function retrieval from
the correlation of passive measurements in a space-dependent
material and a new expression for Green's function retrieval in a time-dependent material.

After an introduction of a matrix-vector wave equation, we discuss propagator matrices  for both types of material. Since the initial condition
for a propagator matrix in a time-dependent material follows from the boundary condition for a propagator matrix in a space-dependent material by interchanging
the time- and space-coordinates, the propagator matrices for both types of material are interrelated in the same way.
This also applies to representations and reciprocity theorems involving propagator matrices, and to Marchenko-type focusing functions.}
\end{summary}


\section{Introduction}\label{sec1}

A wave that encounters a temporal change of material parameters (a so-called time boundary) undergoes reflection and transmission \citep{Xiao2014OL}, 
similar to a wave that is incident on a spatial change of material parameters (a space boundary).
Although research on wave propagation and scattering in time-dependent materials has been around for several decades  \citep{Morgenthaler58IRE, Jiang75IEEE},
recent advances in the construction of dynamic metamaterials have given this field of research a significant boost \citep{Caloz2020IEEE1}.
Whereas most applications concern electromagnetic waves \citep{Mounaix2020NC, Apffel2022PRL, Moussa2023NP}, 
mechanical waves show a similar scattering behaviour when confronted with a temporal change of parameters
\citep{Bacot2016NP, Fink2017EPJ, Peng2020JPC, Hidalgo2023PRL}.
In particular, Fink and coworkers \citep{Bacot2016NP, Fink2017EPJ} show that water waves propagate back to their point of origin when the restoring force 
responsible for wave propagation (gravity), and hence the propagation velocity, is temporarily changed by a vertical acceleration.

Several authors have discussed the analogy between the underlying equations for space-dependent and for time-dependent materials 
\citep{Xiao2014OL, Torrent2018PRB, Mendonca2002PS,  Salem2015arXiv, Caloz2020IEEE2}.
For example, the roles of the time- and space-coordinates in the 1D wave equation for a 
space-dependent material are interchanged in the 1D wave equation for a time-dependent material.
Despite the simple relations between the wave equations, the relation between the solutions of these equations 
(i.e., the wave fields in space-dependent and in time-dependent materials) 
is less straightforward. The reason for this is that the causality conditions are the same in both types of material. 
Only when the initial conditions and boundary conditions would be interchanged
(along with the interchangement of time- and space-coordinates in the wave equations), the solutions would obey a simple relation as well.

The aim of this paper is to discuss a number of fundamental aspects of wave propagation and scattering in space-dependent 
and in time-dependent materials and compare these in a systematic way.
Our discussion partly overlaps with earlier reviews, such as the excellent paper by \cite{Caloz2020IEEE2}, but we also discuss new results.
We use a unified notation for different wave phenomena (electromagnetic, acoustic, elastodynamic), 
so that all relations discussed in this paper hold simultaneously for these phenomena. 
For simplicity, we restrict ourselves to
1D waves only. We discuss Green's functions, propagation invariants, reciprocity theorems, wave field representations and expressions for Green's function retrieval.
In most of these cases, the derived solutions for  space-dependent and time-dependent materials are not exchangeable as a result of non-exchangeable 
causality conditions.
We also discuss propagator matrices for  space-dependent and time-dependent materials and show that they are completely exchangeable as a result of interchangeable
boundary and initial conditions. Finally, we discuss Marchenko-type focusing functions for both types of material and show that they are also exchangeable.

\section{Unified basic equations and constitutive relations for 1D wave fields}

Throughout this paper, we consider 1D wave fields as a function of space (denoted by $x$) and time (denoted by $t$). We take $x$ increasing towards the right.
Using analogies between electromagnetic, acoustic and elastodynamic waves  
\citep{Carcione95WM, Hoop95Book, Wapenaar2001RS, Carcione2002SGG, Hoop2014WM, Burns2020NJP}, the basic equations in a unified notation are
\begin{eqnarray}
\partial_t U + \partial_x Q &=& a,\label{eq1}\\
\partial_t V + \partial_x P &=& b,\label{eq2}
\end{eqnarray}
where $\partial_x$ and $\partial_t$ denote partial derivatives with respect to space and time, respectively,
$U(x,t)$, $V(x,t)$, $P(x,t)$ and $Q(x,t)$ are space- and time-dependent wave-field quantities 
and $a(x,t)$ and $b(x,t)$ are space- and time-dependent source quantities, see Table \ref{table1}.
The wave-field quantities are mutually related via the following constitutive equations 
\begin{eqnarray}
U &=& \alpha P,\label{eq3}\\
V &=& \beta Q,\label{eq4}
\end{eqnarray}
where $\alpha(x,t)$ and $\beta(x,t)$ are the parameters of space- and time-dependent materials, see Table \ref{table1}. 
Rows 1 and 2 contain the quantities for  electromagnetic wave propagation, with TE standing for transverse electric and TM for transverse magnetic. The quantities are
electric flux densities $D_y(x,t)$ and $D_z(x,t)$, magnetic flux densities $B_y(x,t)$ and $B_z(x,t)$,
electric field strengths $E_y(x,t)$ and $E_z(x,t)$, magnetic field strengths $H_y(x,t)$ and $H_z(x,t)$, permittivity $\varepsilon(x,t)$, permeability $\mu(x,t)$,
external electric current densities $J^{\rm e}_y(x,t)$ and $J^{\rm e}_z(x,t)$ and external magnetic current densities $J^{\rm m}_y(x,t)$ and $J^{\rm m}_z(x,t)$. 
The quantities in row 3, associated to  acoustic wave propagation in a  fluid material, are 
 dilatation $\Theta(x,t)$, longitudinal mechanical momentum density $m_x(x,t)$, 
 acoustic pressure $p(x,t)$, longitudinal particle velocity $v_x(x,t)$,  compressibility $\kappa(x,t)$, mass density $\rho(x,t)$, 
volume-injection rate density $q(x,t)$ and external longitudinal force density $f_x(x,t)$.
For  horizontally polarised shear (SH) waves in a  solid material, we have in row 4 
transverse mechanical momentum density $m_y(x,t)$, shear strain $e_{yx}(x,t)$, transverse particle velocity $v_y(x,t)$, shear stress $\tau_{yx}(x,t)$,
mass density $\rho(x,t)$, shear modulus $\mu(x,t)$, external transverse force density $f_y(x,t)$ and external shear deformation rate density $h_{yx}(x,t)$.

\begin{table}
\caption{Quantities in unified equations (\ref{eq1}) -- (\ref{eq4}).}\label{table1}
\begin{center}
\begin{tabular}
{||l||c|c||c|c||c|c||c|c||}
\hline\hline
& $U$ & $V$  &$P$ & $Q$  &$\alpha$ &$\beta$  & $a$ & $b$  \\
\hline\hline
1. TE waves  & $D_y$ & $B_z$ & $E_y$ & $H_z$  &$\varepsilon$ &$\mu$  &  $-J_y^{\rm e}$ & $-J_z^{\rm m}$  \\
\hline
2. TM waves  & $B_y$ & $-D_z$ & $H_y$ & $-E_z$  &$\mu$ &$\varepsilon$  & $-J_y^{\rm m}$ & $J_z^{\rm e}$  \\
\hline
3. Acoustic waves & $-\Theta$ & $m_x$ & $p$ & $v_x$ &$\kappa$ &$\rho$  & $q$ & $f_x$  \\
\hline
4. SH waves & $m_y$ & $-2e_{yx}$ & $v_y$ & $-\tau_{yx}$  &$\rho$ &$\mu^{-1}$ & $f_y$ & $2h_{yx}$ \\
\hline
\hline
\end{tabular}
\end{center}
\end{table}

In the following we consider  either space-dependent parameters 
($\alpha(x)$ and $\beta(x)$) or time-dependent parameters ($\alpha(t)$ and $\beta(t)$). 
For discussions on wave propagation and scattering in materials that are
both space- and time-dependent, we refer to \cite{Apffel2022PRL, Caloz2020IEEE2, Manen2024arXiv};
for non-reciprocal wave propagation due to ``travelling wave modulation'', see 
\cite{Lurie2007Book, Willis2011RS, Willis2012CRM, Trainiti2016NJP, Nassar2017JMPS, Nassar2017RS, Goldsberry2019JASA, Sotoodehfar2023OE}.

\section{Wave equations and Green's functions}\label{sec3}

In this and subsequent sections, the first subsection reviews a specific subject for a space-dependent material. This serves as an introduction to the second subsection,
which discusses the same subject for a time-dependent material, including the analogies and differences.

\subsection{Space-dependent material}\label{sec4.1}

We consider a space-dependent material that is constant over time, with parameters $\alpha(x)$ and $\beta(x)$.
Substituting the constitutive equations (\ref{eq3}) and (\ref{eq4}) 
into the basic equations (\ref{eq1}) and (\ref{eq2}), using the fact that $\alpha(x)$ and $\beta(x)$ are independent of time, gives 
\begin{eqnarray}
\alpha\partial_t P +\partial_xQ &=& a,\label{eq6}\\
\beta\partial_t Q + \partial_x P &=& b.\label{eq7}
\end{eqnarray}
For a space-dependent material with piecewise continuous parameters, these equations are supplemented with boundary conditions at all points where $\alpha(x)$ and $\beta(x)$ 
undergo a finite jump. The boundary conditions are that $P(x,t)$ and $Q(x,t)$ are continuous at those points 
(see \ref{AppA1} for a review of reflection and transmission coefficients at a space boundary).

We obtain a second order wave equation for the field $P(x,t)$ by eliminating $Q(x,t)$ from equations (\ref{eq6}) and (\ref{eq7}), according to
\begin{eqnarray}
 \frac{1}{\beta c^2}\partial_t^2P - \partial_x\Bigl(\frac{1}{\beta}\partial_xP\Bigr) =  \partial_t a - \partial_x\Bigl(\frac{b}{\beta}\Bigr),\label{eq7a}
\end{eqnarray}
with propagation velocity $c(x)$ given by 
\begin{eqnarray}
c=\frac{1}{\sqrt{\alpha\beta}},\label{eq8a}
\end{eqnarray}
with space-dependent parameters $\alpha(x)$ and $\beta(x)$.
We also define 
\begin{eqnarray}
\eta=\sqrt{\frac{\beta}{\alpha}}=\beta c=\frac{1}{\alpha c},\label{eq8b}
\end{eqnarray}
where $\eta$ stands for impedance in the case of TE and acoustic waves (rows 1 and 3 of Table \ref{table1})
or admittance in the case of TM and SH waves (rows 2 and 4 of Table \ref{table1}).

We define the Green's function ${\cal G}_x(x,x_0,t)$ as the response to an impulsive point source 
$\delta(x-x_0)\delta(t)$, hence
\begin{eqnarray}
\frac{1}{\beta c^2}\partial_t^2{\cal G}_x - \partial_x\Bigl(\frac{1}{\beta}\partial_x{\cal G}_x\Bigr)  =   \delta(x-x_0) \delta(t),\label{eq9a}
\end{eqnarray}
with causality condition 
\begin{eqnarray}
{\cal G}_x(x,x_0,t)=0\quad \mbox{for}\quad t<0.\label{eqcaus1}
\end{eqnarray}
This condition implies that ${\cal G}_x(x,x_0,t)$ is outward propagating for $|x|\to\infty$. 
The subscript $x$ in ${\cal G}_x$ denotes that this is the Green's function of a space-dependent material. 

A simple representation for $P(x,t)$ is obtained when $P$ and ${\cal G}_x$ are defined in the same material and both are outward propagating for $|x|\to\infty$.
Whereas $P(x,t)$ is the response to an arbitrary source distribution $\partial_ta(x,t)$ (equation (\ref{eq7a}), assuming $b=0$), ${\cal G}_x(x,x_0,t)$ is the response to an impulsive 
point source at an arbitrary location $x_0$ at $t=0$ (equation (\ref{eq9a})).
Because equations (\ref{eq7a}) and (\ref{eq9a}) are linear, a representation for $P(x,t)$ follows by applying 
Huygens' superposition principle. Assuming $\partial_ta(x,t)$ is causal, i.e., $\partial_ta(x,t)=0$ for $t<0$, this gives
\citep{Morse53Book, Bleistein84Book}
\begin{eqnarray}
P(x,t)=\int_{-\infty}^\infty{\rm d}x' \int_0^t {\cal G}_x(x,x',t-t')\partial_{t'}a(x',t'){\rm d}t'.\label{eq31k}
\end{eqnarray}
This representation is a special case of the more general representation for a space-dependent material, derived in section \ref{sec7.1}.

\begin{figure}
\centerline{\epsfysize=7. cm \epsfbox{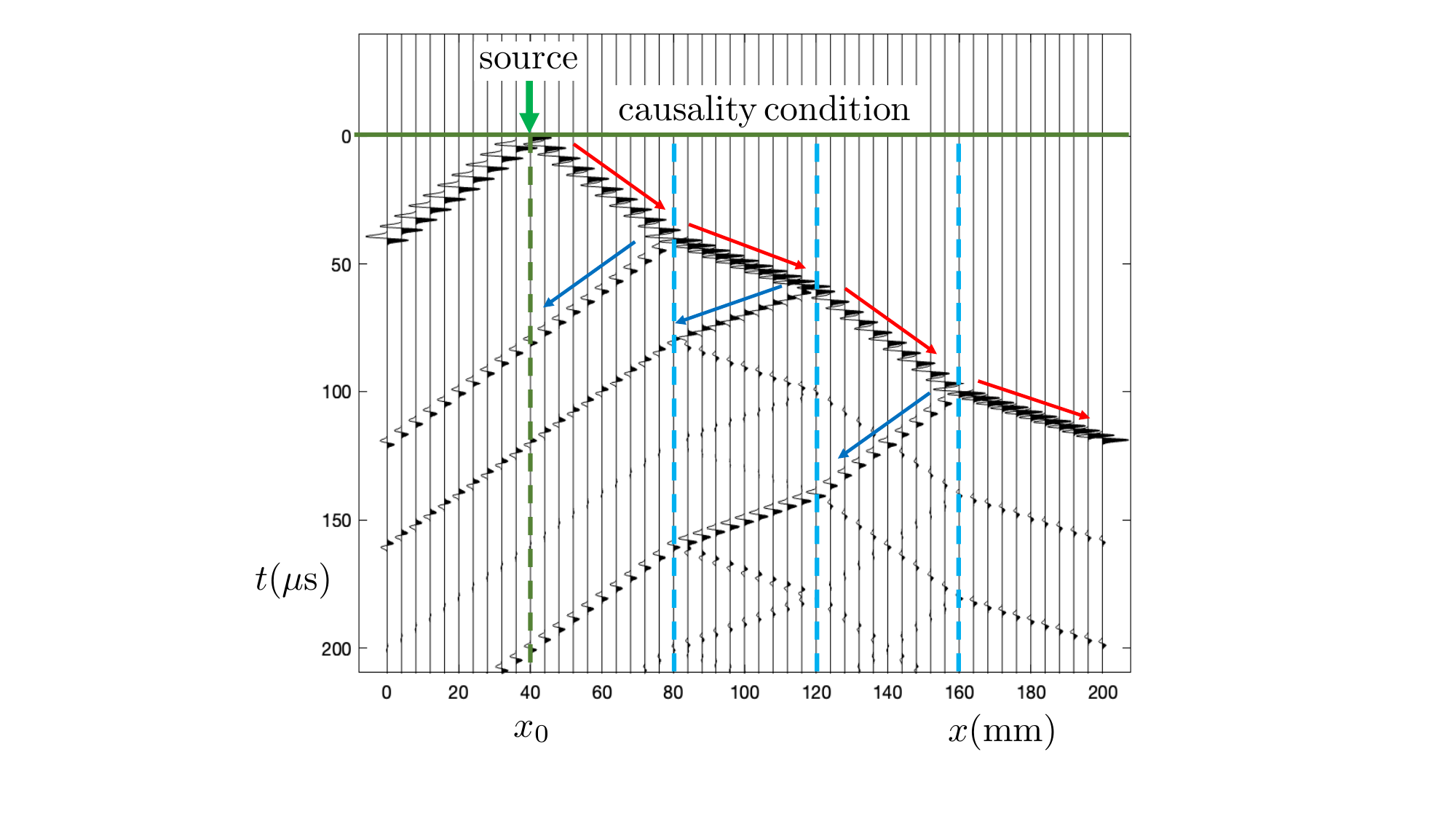}}
\caption{Green's function ${\cal G}_x(x,x_0,t)$ (convolved with a temporal wavelet) for a piecewise homogeneous space-dependent material.}\label{Fig1}
\end{figure}

We discuss a numerical example of an acoustic Green's function for a piecewise homogeneous material, consisting of five homogeneous slabs, each with a thickness of 40 mm.
The propagation velocities are 1.0, 1.0, 2.0, 1.0 and 2.2 km/s, respectively.
The half-spaces to the left and the right of the space-dependent material are homogeneous,
with the same velocities as the first and last slab, respectively. The parameter $\beta$ is constant throughout.
The source is located between the first and the second slab, at $x_0=40$ mm. 
We use a recursive ``layer-code'' method \citep{Kennett83Book} to model the response to this source.
Figure \ref{Fig1} shows an $x,t$-diagram of ${\cal G}_x(x,x_0,t)$, convolved in time with a temporal wavelet with a central frequency $\omega_0/2\pi=300$ kHz.
The causality condition of equation (\ref{eqcaus1}) implies that the Green's function  is zero above the green line at $t=0$. 
The red arrows indicate the rightward propagating primary wave and the blue arrows the leftward propagating primary reflections. 
Multiply scattered waves are also clearly visible. Note that the field is outward propagating for $x=0$ and $x=200$ mm (and hence for $|x|\to\infty$, 
since the left and right half-spaces are homogeneous).

\subsection{Time-dependent material}\label{sec4.2}

We consider a  time-dependent homogeneous material with parameters $\alpha(t)$ and $\beta(t)$.
Substituting the constitutive equations (\ref{eq3}) and (\ref{eq4}) 
into the basic equations (\ref{eq1}) and (\ref{eq2}), using the fact that $\alpha(t)$ and $\beta(t)$ are independent of space, gives 
\begin{eqnarray}
\partial_t U + \frac{1}{\beta}\partial_x V &=& a,\label{eq49}\\
\partial_t V + \frac{1}{\alpha}\partial_x U &=& b.\label{eq50}
\end{eqnarray}
For a time-dependent material with piecewise continuous parameters, these equations are supplemented with boundary conditions at all time instants 
where $\alpha(t)$ and $\beta(t)$ undergo a finite jump. The boundary conditions are that $U(x,t)$ and $V(x,t)$ are continuous at those time instants
(see \ref{AppA2} for a review of reflection and transmission coefficients at a time boundary).

We obtain a second order wave equation for the field $U(x,t)$ by eliminating $V(x,t)$  from equations (\ref{eq49}) and (\ref{eq50}), according to
\begin{eqnarray}
\partial_t(\beta\partial_tU) - \beta c^2\partial_x^2U =   \partial_t(\beta a) - \partial_x b,\label{eq31a}
\end{eqnarray}
with propagation velocity $c(t)$ again given by equation (\ref{eq8a}), this time with time-dependent parameters $\alpha(t)$ and $\beta(t)$.

Note that equations (\ref{eq6}), (\ref{eq7}) and (\ref{eq7a}) can be transformed into equations (\ref{eq50}), (\ref{eq49}) and (\ref{eq31a}) and vice-versa, 
by the following mapping
\begin{eqnarray}
\{P,Q,x,t,\alpha,\beta,a,b,c\}\leftrightarrow \{U,V,t,x,\alpha^{-1},\beta^{-1},b,a,c^{-1}\}.\label{eqmap}
\end{eqnarray}

We define the Green's function ${\cal G}_t(x,t,t_0)$ as the response to an impulsive point source 
$\delta(x)\delta(t-t_0)$, hence
\begin{eqnarray}
\partial_t(\beta\partial_t{\cal G}_t) - \beta c^2\partial_x^2{\cal G}_t =  \delta(x)\delta(t-t_0),\label{eq36b}
\end{eqnarray}
with causality condition
\begin{eqnarray}
{\cal G}_t(x,t,t_0)=0\quad \mbox{for}\quad t<t_0.\label{eqcaus2}
\end{eqnarray}
This condition implies that ${\cal G}_t(x,t,t_0)$ is outward propagating for $|x|\to\infty$. 
The subscript $t$ in ${\cal G}_t$ denotes that this is the Green's function of a time-dependent material. 

A simple representation for $U(x,t)$ is obtained when $U$ and ${\cal G}_t$ are defined in the same material and both are outward propagating for $|x|\to\infty$.
Whereas $U(x,t)$ is the response to a source distribution $-\partial_xb(x,t)$ (equation (\ref{eq31a}), assuming $a=0$), ${\cal G}_t(x,t,t_0)$ is the response to an impulsive 
point source at $x=0$ at an arbitrary time $t_0$ (equation (\ref{eq36b})).
Because equations (\ref{eq31a}) and (\ref{eq36b}) are linear, a representation for $U(x,t)$ follows by applying 
Huygens' superposition principle. Assuming $\partial_xb(x,t)$ is causal, i.e., $\partial_xb(x,t)=0$ for $t<0$, this gives
\begin{eqnarray}
U(x,t)=-\int_{-\infty}^\infty{\rm d}x' \int_0^t {\cal G}_t(x-x',t,t')\partial_{x'}b(x',t'){\rm d}t'.\label{eq38k}
\end{eqnarray}
Using equation (\ref{eq3}), a representation for $P(x,t)$ follows from $P(x,t)=\frac{1}{\alpha(t)}U(x,t)$, with $U(x,t)$ given by equation (\ref{eq38k}).
The representation of equation (\ref{eq38k})  is a special case of the more general representation for a time-dependent material, derived in section \ref{sec7.2}.

\begin{figure}
\centerline{\epsfysize=7. cm \epsfbox{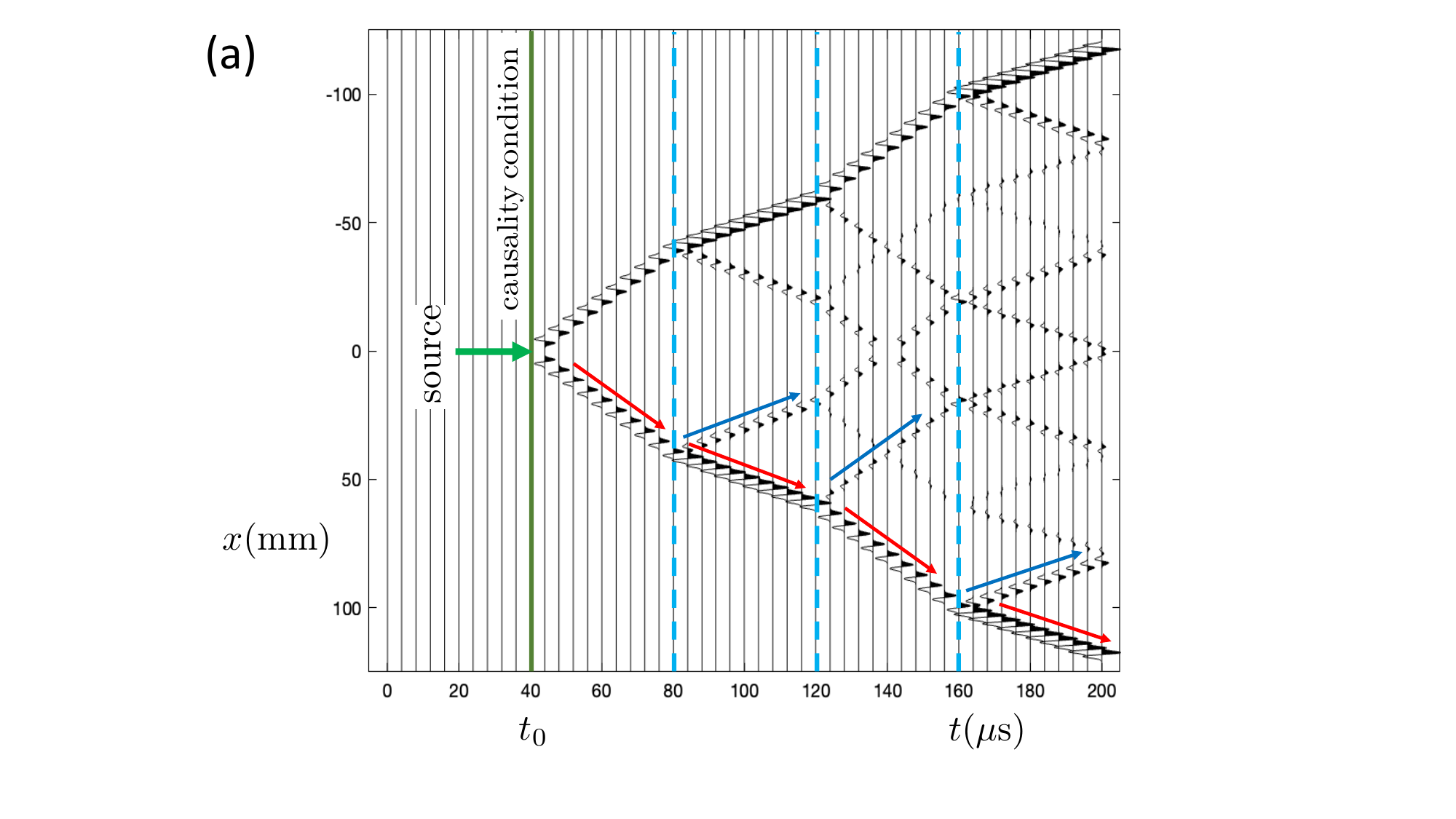}}
\centerline{\epsfysize=7. cm \epsfbox{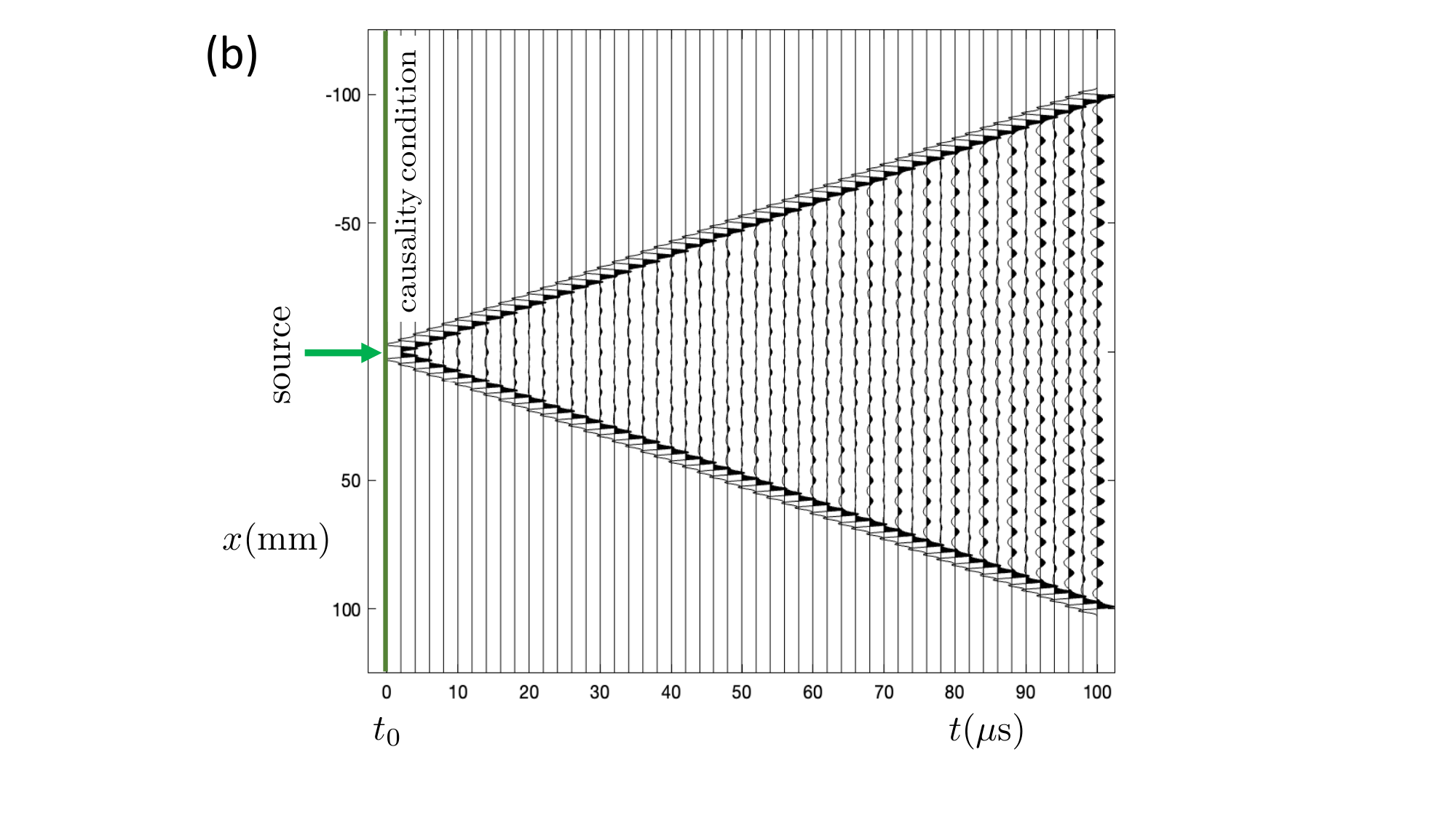}}
\caption{Green's function ${\cal G}_t(x,t,t_0)$ (convolved with a spatial wavelet) for (a) a piecewise constant and (b) a sinusoidally modulated 
time-dependent material.}\label{Fig2}
\end{figure}

We discuss a numerical example of an acoustic Green's function for a piecewise constant material, consisting of five time-independent slabs.
Following the mapping of equation (\ref{eqmap}), we ``construct'' this material from the material used for the numerical example in section \ref{sec4.1}, 
with time and space interchanged and with the reciprocal propagation velocities.
For convenience, we define 1 km as the unit of distance and 1 km/s as the unit of velocity.
With this definition, the reciprocal propagation velocities are 1.0, 1.0, 0.5, 1.0 and 0.455 km/s, respectively.
The half-times before and after the time-dependent material are constant, with the same velocities as the first and last slab, respectively.  
 The parameter $\beta$ is again constant throughout.
Note that 1 mm, which is actually 1 $\mu$km, maps to 1 $\mu$s and vice-versa. 
Hence, the slab thickness of 40 mm is mapped to a slab duration of 40 $\mu$s.
The source is located between the first and second slab, at $t_0=40\, \mu$s.
Figure \ref{Fig2}a shows an $x,t$-diagram of ${\cal G}_t(x,t,t_0)$, convolved in space with a spatial wavelet with a central wavenumber $k_0/2\pi=300*10^3$ km$^{-1}$.
The causality condition of equation (\ref{eqcaus2}) implies that the Green's function  is zero left of the green line at $t=t_0$. 
The red arrows indicate the rightward propagating primary wave (i.e., in the $+x$ direction) and the blue arrows the leftward propagating primary reflections
(in the $-x$ direction). 
Multiply scattered waves are also clearly visible.

Since the causality conditions (equations (\ref{eqcaus1}) and (\ref{eqcaus2})) do not follow 
the mapping of equation (\ref{eqmap}), the  $x,t$-diagrams of the Green's functions
for space-dependent and time-dependent materials (Figures \ref{Fig1} and \ref{Fig2}a) are very different. 
For the specially designed case considered here (with reciprocal velocities), 
only the rightward propagating primary waves (indicated by the red arrows) exhibit interchangeable kinematical behaviour between the two cases 
(but they have different amplitudes). All other events are different in these figures.
 Whereas the multiply scattered waves in ${\cal G}_x(x,x_0,t)$ in Figure \ref{Fig1} 
consist of ongoing reverberations between space boundaries, 
the multiply scattered waves in ${\cal G}_t(x,t,t_0)$ in Figure \ref{Fig2}a are the result of ``forward-in-time'' reflections 
and transmissions at time boundaries (see also Figure \ref{Fig0} in the Appendix); their total number is finite.
In section \ref{sec10} we discuss propagator matrices for space-dependent and time-dependent materials 
and show that these follow the mapping of equation (\ref{eqmap}) for all events. 
Moreover, in section \ref{sec10.2} we show how ${\cal G}_t(x,t,t_0)$ is related to one of the elements of the propagator matrix for a time-dependent material.

Finally, Figure \ref{Fig2}b is an example of an acoustic Green's function ${\cal G}_t(x,t,t_0)$ (convolved with the same spatial wavelet as in Figure \ref{Fig2}a) 
of a sinusoidally modulated time-dependent material, with propagation velocity 
$c(t)=1+\frac{1}{10}\sin(\pi t/2)$ km/s (with $t$ in $\mu$s, ranging from $t_0=0\, \mu$s to $t_N=100\, \mu$s) and  constant $\beta$. 
For the modeling we divided the velocity profile into 4000 constant velocity slabs with a duration of  0.025 $\mu$s each. 
Note the complex scattering behaviour and the increasing amplitudes of the scattered events with time.
To compensate for the increasing amplitudes, \cite{Torrent2018PRB} introduce a dissipative time-dependent material. 

\section{Propagation invariants}

\subsection{Space-dependent material}

We review propagation invariants for a space-dependent material  with parameters $\alpha(x)$ and $\beta(x)$.
Given a space- and time-dependent function $P(x,t)$, we define its temporal Fourier transform as
\begin{eqnarray}
\hat P(x,\omega) = \int_{-\infty}^\infty P(x,t)\exp \{i\omega t\}{\rm d}t,\label{eq10}
\end{eqnarray}
with $\omega$ denoting the angular frequency and $i$ the imaginary unit.
With this definition, derivatives with respect to time transform to multiplications with $-i\omega$. Hence,
equations (\ref{eq6}) and (\ref{eq7}) transform to
\begin{eqnarray}
-i\omega\alpha\hat P +\partial_x\hat Q &=&\hat a,\label{eq6f}\\
-i\omega\beta\hat Q + \partial_x\hat P &=&\hat b.\label{eq7f}
\end{eqnarray}
In the following we consider two independent states, indicated by subscripts $A$ and $B$, 
obeying equations (\ref{eq6f}) and (\ref{eq7f}).
In the most general case, sources, material parameters and wave fields may be different in the two states.  
We derive relations between these states. First we consider the quantity $\partial_x\{\hat P_A\hat Q_B - \hat Q_A\hat P_B\}$.
Applying the product rule for differentiation, using equations (\ref{eq6f}) and (\ref{eq7f}) for states $A$ and $B$
to get rid of the derivatives,  we obtain
\begin{eqnarray}
\partial_x\{\hat P_A\hat Q_B - \hat Q_A\hat P_B\} &=&i\omega(\alpha_B-\alpha_A)\hat P_A\hat P_B-i\omega(\beta_B-\beta_A)\hat Q_A\hat Q_B\nonumber\\
&&-\hat a_A\hat P_B+\hat b_A\hat Q_B+\hat P_A\hat a_B-\hat Q_A\hat b_B.\label{eq1013}
\end{eqnarray}
This is the local reciprocity theorem of the time-convolution type \citep{Hoop95Book, Fokkema93Book}, in which products like $\hat P_A\hat Q_B$ 
correspond to convolutions along the time coordinate in the $x,t$-domain.
Next, we consider the quantity $\partial_x\{\hat P_A^*\hat Q_B + \hat Q_A^*\hat P_B\}$ (where the asterisk denotes complex conjugation) and apply the same operations, yielding
\begin{eqnarray}
\partial_x\{\hat P_A^*\hat Q_B + \hat Q_A^*\hat P_B\} &=&i\omega(\alpha_B-\alpha_A)\hat P_A^*\hat P_B+i\omega(\beta_B-\beta_A)\hat Q_A^*\hat Q_B\nonumber\\ 
&&+\hat a_A^*\hat P_B+\hat b_A^*\hat Q_B+\hat P_A^*\hat a_B+\hat Q_A^*\hat b_B.\label{eq1014}
\end{eqnarray}
This is the local reciprocity theorem of the time-correlation type \citep{Hoop95Book, Bojarski83JASA}, in which products like $\hat P_A^*\hat Q_B$ 
correspond to correlations along the time coordinate in the $x,t$-domain. In section \ref{sec6.1} we will use equations (\ref{eq1013}) and (\ref{eq1014}) as the basis 
for deriving global reciprocity theorems of the time-convolution and time-correlation type. 
Here we use these equations to derive propagation invariants for a space-dependent material.
To this end we take identical material parameters in states $A$ and $B$ and we assume that there are no sources. With this, equations (\ref{eq1013}) and (\ref{eq1014})
simplify to
\begin{eqnarray}
\partial_x\{\hat P_A\hat Q_B - \hat Q_A\hat P_B\} &=& 0,\label{eq1015}\\
\partial_x\{\hat P_A^*\hat Q_B + \hat Q_A^*\hat P_B\} &=& 0,\label{eq1016}
\end{eqnarray}
hence, $\hat P_A\hat Q_B - \hat Q_A\hat P_B$ and $\hat P_A^*\hat Q_B + \hat Q_A^*\hat P_B$ are space-propagation invariants, i.e., 
they are independent of the space coordinate $x$.
This holds for continuously varying material parameters $\alpha(x)$ and $\beta(x)$. 
For a material with piecewise continuous parameters, the boundary conditions state that $\hat P$ and $\hat Q$ are continuous at all points where
$\alpha(x)$ and $\beta(x)$ are discontinuous. This implies that the space-propagation invariants also hold for  a space-dependent material with piecewise continuous parameters.
Propagation invariants find applications in the analysis of symmetry properties of reflection and transmission responses and have been used for the design of efficient 
numerical modelling schemes \citep{Haines88GJI, Kennett90GJI, Koketsu91GJI, Takenaka93WM}.

For the special case that the wave fields in states $A$ and $B$ are identical, we may drop the subscripts $A$ and $B$.
The first space-propagation invariant then vanishes and is no longer useful. The second space-propagation invariant simplifies to $2\Re\{\hat P^*\hat Q\}$, 
where $\Re$ denotes the real part.
We define $\hat\j(x,\omega)=\frac{1}{2}\Re\{\hat P^*\hat Q\}$ as the net power-flux density in the $x$-direction in the $x,\omega$-domain.
Hence, the net power-flux density is conserved (i.e., it is independent of $x$)  in a space-dependent material with piecewise continuous parameters.

\subsection{Time-dependent material}

We derive propagation invariants for a time-dependent material  with parameters $\alpha(t)$ and $\beta(t)$.
Given a space- and time-dependent function $U(x,t)$, we define its spatial Fourier transform as
\begin{eqnarray}
\check U(k,t) = \int_{-\infty}^\infty U(x,t)\exp \{-ikx\}{\rm d}x,\label{eq60}
\end{eqnarray}
with $k$ denoting the wavenumber.
Following common conventions, we use opposite signs in the exponentials in the temporal and spatial Fourier transforms (equations (\ref{eq10}) and (\ref{eq60})).
With this definition, derivatives with respect to space transform to multiplications with $ik$. Hence,
equations (\ref{eq49}) and (\ref{eq50}) transform to
\begin{eqnarray}
\partial_t \check U + \frac{ik}{\beta} \check V &=& \check a,\label{eq49f}\\
\partial_t \check V + \frac{ik}{\alpha} \check U &=& \check b.\label{eq50f}
\end{eqnarray}

Note that equations (\ref{eq6f}) and (\ref{eq7f})  can be transformed into equations (\ref{eq50f}) and (\ref{eq49f}) and vice-versa, 
by the following modified mapping
\begin{eqnarray}
\{\hat P,\hat Q,x,\omega,\alpha,\beta,\hat a,\hat b,c\}\leftrightarrow \{\check U,\check V,t,-k,\alpha^{-1},\beta^{-1},\check b,\check a,c^{-1}\}. \label{eqmap2}
\end{eqnarray}
In the following we consider two independent states, indicated by subscripts $A$ and $B$, 
obeying equations (\ref{eq49f}) and (\ref{eq50f}).
In the most general case, sources, material parameters and wave fields may be different in the two states.  
Applying the mapping of equation (\ref{eqmap2}) 
to equations (\ref{eq1013}) and (\ref{eq1014}) yields the local reciprocity theorems of the space-convolution and space-correlation type, in which
products like $\check U_A\check V_B$ and $\check U_A^*\check V_B$ correspond to convolutions and correlations, respectively, along the space coordinate in the $x,t$-domain.
In section \ref{sec6.2} we derive
global reciprocity theorems of the space-convolution and space-correlation type.
Here we  derive propagation invariants for a time-dependent material.
To this end we take identical material parameters in states $A$ and $B$ and we assume that there are no sources.
Applying the mapping of equation (\ref{eqmap2}) to equations (\ref{eq1015}) and (\ref{eq1016}) yields
\begin{eqnarray}
\partial_t\{\check U_A\check V_B - \check V_A\check U_B\} &=& 0,\label{eq1015t}\\
\partial_t\{\check U_A^*\check V_B + \check V_A^*\check U_B\} &=& 0,\label{eq1016t}
\end{eqnarray}
hence, $\check U_A\check V_B - \check V_A\check U_B$ and $\check U_A^*\check V_B + \check V_A^*\check U_B$ are time-propagation invariants, i.e.,
 they are independent of the time coordinate $t$.
This holds for continuously varying material parameters $\alpha(t)$ and $\beta(t)$. 
For a material with piecewise continuous parameters, the boundary conditions state that $\check U$ and $\check V$ are continuous at all time instants where
$\alpha(t)$ and $\beta(t)$ are discontinuous. This implies that the time-propagation invariants also hold for  a time-dependent material with piecewise continuous parameters.

For the special case that the wave fields in states $A$ and $B$ are identical, we may drop the subscripts $A$ and $B$.
The first time-propagation invariant then vanishes and is no longer useful. The second time-propagation invariant simplifies to $2\Re\{\check U^*\check V\}$.
We define $\check M(k,t)=\frac{1}{2}\Re\{\check U^*\check V\}$ as the net field-momentum density \citep{Burns2020NJP, Feynmann63Book2}
 in the $x$-direction in the $k,t$-domain
(to be distinguished from the mechanical momentum densities $m_x$ and $m_y$ in Table \ref{table1}).
Hence, the net field-momentum density is conserved (i.e., it is independent of $t$) in a time-dependent material with piecewise continuous parameters. 

Using equations (\ref{eq3}), (\ref{eq4}) and (\ref{eq8a}), we obtain for the net power-flux density $\check \j(k,t)$ in the $x$-direction  in the $k,t$-domain, 
defined as $\frac{1}{2}\Re\{\check P^*\check Q\}$, 
\begin{eqnarray}
\check \j(k,t)=c^2(t)\check M(k,t).\label{eq51uj}
\end{eqnarray}
Hence, whereas the net field-momentum density $\check M(k,t)$ is conserved, 
the net power-flux density $\check \j(k,t)$ is not conserved (i.e., it is dependent on $t$) in a time-dependent material.
This is explained as the result of energy being added to or extracted from the wave field by the external source that modulates the material parameters 
\citep{Morgenthaler58IRE, Mendonca2002PS, Caloz2020IEEE2}.

\section{Reciprocity theorems}

\subsection{Space-dependent material}\label{sec6.1}

We review general reciprocity theorems for a space-dependent material  with piecewise continuous parameters $\alpha(x)$ and $\beta(x)$.
Integrating both sides of equations (\ref{eq1013}) and (\ref{eq1014})  from $x_b$ to $x_e$ (with subscripts $b$ and $e$ standing for ``begin'' and ``end''),
taking into account that $\hat P$ and $\hat Q$ are continuous at points where $\alpha(x)$ and $\beta(x)$ are discontinuous, yields
\begin{eqnarray}
\hspace{-.7cm}&&\{\hat P_A\hat Q_B - \hat Q_A\hat P_B\}\Bigr|_{x_b}^{x_e} = 
\int_{x_b}^{x_e}\Bigl(
i\omega(\alpha_B-\alpha_A)\hat P_A\hat P_B-i\omega(\beta_B-\beta_A)\hat Q_A\hat Q_B\nonumber\\&&\hspace{5cm}
-\hat a_A\hat P_B+\hat b_A\hat Q_B+\hat P_A\hat a_B-\hat Q_A\hat b_B\Bigr){\rm d}x,\label{eq1017}\\
\hspace{-.7cm}&&\{\hat P_A^*\hat Q_B + \hat Q_A^*\hat P_B\}\Bigr|_{x_b}^{x_e} = 
\int_{x_b}^{x_e}\Bigl(
i\omega(\alpha_B-\alpha_A)\hat P_A^*\hat P_B+i\omega(\beta_B-\beta_A)\hat Q_A^*\hat Q_B\nonumber\\&&\hspace{5cm}
+\hat a_A^*\hat P_B+\hat b_A^*\hat Q_B+\hat P_A^*\hat a_B+\hat Q_A^*\hat b_B\Bigr){\rm d}x.\label{eq1017c}
\end{eqnarray}
These are the global reciprocity theorems of the time-convolution and time-correlation type, respectively, for a space-dependent material 
\citep{Hoop95Book, Fokkema93Book, Bojarski83JASA, Rayleigh78Book, Lorentz1895KNAW}.
We use equation (\ref{eq1017}) in section \ref{sec7.1} to derive a general wave field representation and in section \ref{sec8.1}  we use equation (\ref{eq1017c})
to derive an expression for Green's function retrieval, both for space-dependent materials.
Here we discuss two special cases of equations (\ref{eq1017}) and (\ref{eq1017c}).

First we derive an expression for source-receiver reciprocity of the Green's function of a space-dependent material from equation (\ref{eq1017}).
To this end we take identical material parameters in both states, i.e., $\alpha_A=\alpha_B=\alpha$ and  $\beta_A=\beta_B=\beta$ and
we assume that the material is homogeneous for $x\le x_b$ and for $x\ge x_e$. For state $A$ 
we take a Green's state with a unit source at $x_A$ between $x_b$ and $x_e$, hence, we substitute  
 $-i\omega\hat a_A(x,\omega)=\delta(x-x_A)$, $\hat b_A(x,\omega)=0$, $\hat P_A(x,\omega)=\hat {\cal G}_x(x,x_A,\omega)$ and, 
 using equation (\ref{eq7f}), $\hat Q_A(x,\omega)=\frac{1}{i\omega\beta(x)}\partial_x\hat {\cal G}_x(x,x_A,\omega)$.
For state $B$ we take a Green's state with a unit source at $x_B$ between $x_b$ and $x_e$, and we substitute similar expressions.
 At $x_b$ and $x_e$ the field is 
 leftward and rightward propagating, respectively (see for example Figure \ref{Fig1}, with $x_b=0$ and $x_e=200$ mm),
 i.e., proportional to $\exp(-i\omega x/c(x_b))$ and $\exp(i\omega x/c(x_e))$, respectively.
 Hence
 \begin{eqnarray}
 \hat Q_A(x_b,\omega) &=&\frac{1}{i\omega\beta(x_b)}\partial_x\hat {\cal G}_x(x,x_A,\omega)|_{x=x_b}=-\frac{1}{\eta(x_b)}\hat {\cal G}_x(x_b,x_A,\omega),\label{eq54a}\\
 \hat Q_A(x_e,\omega) &=&\frac{1}{i\omega\beta(x_e)}\partial_x\hat {\cal G}_x(x,x_A,\omega)|_{x=x_e}=+\frac{1}{\eta(x_e)}\hat {\cal G}_x(x_e,x_A,\omega),\label{eq55a}
 \end{eqnarray}
with $\eta$ defined in equation (\ref{eq8b}), and similar expressions for state $B$.
With this, the left-hand side of equation (\ref{eq1017}) vanishes. From the remaining terms on the right-hand side we obtain
\begin{eqnarray}
\hat {\cal G}_x(x_B,x_A,\omega)=\hat {\cal G}_x(x_A,x_B,\omega),\label{eq1022}
\end{eqnarray}
or, in the space-time domain, 
\begin{eqnarray}
{\cal G}_x(x_B,x_A,t)={\cal G}_x(x_A,x_B,t),\label{eq1022t}
\end{eqnarray}
which formulates the classical source-receiver reciprocity relation for a space-dependent material \citep{Hoop95Book, Morse53Book, Hoenders79Optik}. 
These expressions remain valid for arbitrary  $x_A$ and $x_B$  when taking $x_b\to-\infty$ and $x_e\to\infty$.

Second, we derive a power balance for a space-dependent material from equation (\ref{eq1017c}). Taking identical states $A$ and $B$ (hence, identical sources,
material parameters and wave fields), we drop the subscripts $A$ and $B$. Equation (\ref{eq1017c}) thus yields \citep{Hoop95Book, Fokkema93Book}
\begin{eqnarray}
\hat\j(x,\omega)\Bigr|_{x_b}^{x_e} = 
\frac12\Re\int_{x_b}^{x_e}\bigl(\hat a^*\hat P+\hat b^*\hat Q\bigr){\rm d}x.\label{eq1017d}
\end{eqnarray}
This equation states that the power generated by sources in the region between $x_b$ and $x_e$ (the right-hand side) 
is equal to the power leaving this region (the left-hand side). Hence, this equation formulates the power balance for a space-dependent material.

\subsection{Time-dependent material}\label{sec6.2}

We derive general reciprocity theorems for a time-dependent material  with piecewise continuous parameters $\alpha(t)$ and $\beta(t)$.
Applying the mapping of equation (\ref{eqmap2}) to equations (\ref{eq1017}) and (\ref{eq1017c}) yields
\begin{eqnarray}
\hspace{-.7cm}&&\{\check U_A\check V_B - \check V_A\check U_B\}\Bigr|_{t_b}^{t_e} = 
\int_{t_b}^{t_e}\Bigl(
-ik(\alpha_B^{-1}-\alpha_A^{-1})\check U_A\check U_B+ik(\beta_B^{-1}-\beta_A^{-1})\check V_A\check V_B\nonumber\\&&\hspace{5cm}
-\check b_A\check U_B+\check a_A\check V_B+\check U_A\check b_B-\check V_A\check a_B\Bigr){\rm d}t,\label{eq1017t}\\
\hspace{-.7cm}&&\{\check U_A^*\check V_B + \check V_A^*\check U_B\}\Bigr|_{t_b}^{t_e} = 
\int_{t_b}^{t_e}\Bigl(
-ik(\alpha_B^{-1}-\alpha_A^{-1})\check U_A^*\check U_B-ik(\beta_B^{-1}-\beta_A^{-1})\check V_A^*\check V_B\nonumber\\&&\hspace{5cm}
+\check b_A^*\check U_B+\check a_A^*\check V_B+\check U_A^*\check b_B+\check V_A^*\check a_B\Bigr){\rm d}t.\label{eq1017tc}
\end{eqnarray}
These are the global reciprocity theorems of the space-convolution and space-correlation type, respectively, for a time-dependent material.
We use equation (\ref{eq1017t}) in section \ref{sec7.2} to derive a general wave field representation and in section \ref{sec8.2}  we use equation (\ref{eq1017tc})
to derive an expression for Green's function retrieval, both for time-dependent materials.
Here we discuss two special cases of equations (\ref{eq1017t}) and (\ref{eq1017tc}).

First we derive an expression for source-receiver reciprocity of the Green's function of a time-dependent material from equation (\ref{eq1017t}).
Since the causality conditions for the Green's functions do not obey the mapping of equation (\ref{eqmap}), 
the derivation of source-receiver reciprocity is different from that in section \ref{sec6.1}. In particular, for the Green's function of a time-dependent material
there is not an equivalent of leftward and rightward propagating waves at $t_b$ and $t_e$, respectively 
(see for example Figure \ref{Fig2}a, with $t_b=0$ and $t_e=200\, \mu$s).
We take again $\alpha_A=\alpha_B=\alpha$ and  $\beta_A=\beta_B=\beta$.
For state $A$ we take a Green's state with 
an impulse at $t_A$ between $t_b$ and $t_e$, according to $-ik\check b_A(k,t)=\delta(t-t_A)$ and $\check a_A(k,t)=0$.
However, we define $\check U_A(k,t)=\check {\cal G}_t^a(k,t,t_A)$, where $\check {\cal G}_t^a(k,t,t_A)$ is the acausal Green's function, i.e., $\check {\cal G}_t^a(k,t,t_A)=0$ for $t>t_A$
(hence, the impulse at $t_A$ is actually a sink).
Using equation (\ref{eq49f}) we have $\check V_A(k,t)=-\frac{\beta(t)}{ik}\partial_t\check {\cal G}_t^a(k,t,t_A)$.
For state $B$ we take a Green's state with 
an impulsive source at $t_B$ between $t_b$ and $t_e$, according to $-ik\check b_B(k,t)=\delta(t-t_B)$ and $\check a_B(k,t)=0$.
We define $\check U_B(k,t)=\check {\cal G}_t(k,t,t_B)$, 
where $\check {\cal G}_t(k,t,t_B)$ is the causal Green's function, i.e., $\check {\cal G}_t(k,t,t_B)=0$ for $t<t_B$. 
Moreover, $\check V_B(k,t)=-\frac{\beta(t)}{ik}\partial_t\check {\cal G}_t(k,t,t_B)$.
With these definitions, the acausal Green's function is zero for $t=t_e$ and the causal Green's function is zero for $t=t_b$ 
(the latter is seen for example in Figure \ref{Fig2}a, with $t_b=0$).
With this, the left-hand side of equation (\ref{eq1017t}) vanishes. From the remaining terms on the right-hand side we obtain
\begin{eqnarray}
\check {\cal G}_t^a(k,t_B,t_A)=\check {\cal G}_t(k,t_A,t_B),\label{eq1022ta}
\end{eqnarray}
or, in the space-time domain,
\begin{eqnarray}
{\cal G}_t^a(x,t_B,t_A)={\cal G}_t(x,t_A,t_B),\label{eq1022s}
\end{eqnarray}
which formulates source-receiver reciprocity for a time-dependent material.
These expressions remain valid for arbitrary  $t_A$ and $t_B$ when taking $t_b\to-\infty$ and $t_e\to\infty$.
Note that for $t_A<t_B$ these expressions reduce to the trivial relation $0=0$.

Second, we derive a field-momentum balance for a time-dependent material from equation (\ref{eq1017tc}). Taking identical states $A$ and $B$, 
we drop the subscripts $A$ and $B$. Equation (\ref{eq1017tc}) thus yields
\begin{eqnarray}
\check M(k,t)\Bigr|_{t_b}^{t_e} = 
\frac{1}{2}\Re\int_{t_b}^{t_e}\bigl(\check b^*\check U+\check a^*\check V\bigr){\rm d}t.\label{eq1017td}
\end{eqnarray}
This equation states that the field momentum generated by sources in the interval between $t_b$ and $t_e$ (the right-hand side) is equal to the
field momentum leaving this interval (the left-hand side). Hence, this equation formulates the field-momentum balance for a time-dependent material.

\section{Wave field representations}

\subsection{Space-dependent material}\label{sec7.1}

We review a general wave field representation for a space-dependent material  with piecewise continuous parameters $\alpha(x)$ and $\beta(x)$.
Our starting point is the global reciprocity theorem of the time-convolution type for a space-dependent material, formulated by equation (\ref{eq1017}).
For  state $A$ we take the Green's state, hence, we substitute $-i\omega\hat a_A(x,\omega)=\delta(x-x_A)$, $\hat b_A(x,\omega)=0$,
$\hat P_A(x,\omega)=\hat {\cal G}_x(x,x_A,\omega)$ and $\hat Q_A(x,\omega)=\frac{1}{i\omega\beta_A(x)}\partial_x\hat {\cal G}_x(x,x_A,\omega)$.
For state $B$ we take the actual field and drop the subscripts $B$. 
Substitution into equation (\ref{eq1017}), using reciprocity relation (\ref{eq1022}), this yields the following classical representation
\begin{eqnarray}
&&\hspace{-.7cm}\chi(x_A)\hat P(x_A,\omega)=\int_{x_b}^{x_e}\Bigl(-i\omega\hat {\cal G}_x(x_A,x,\omega)\hat a(x,\omega)+
\frac{1}{\beta_A(x)}\{\partial_x\hat {\cal G}_x(x_A,x,\omega)\}\hat b(x,\omega)\Bigr){\rm d}x
\nonumber\\
&&\hspace{-.7cm}+\int_{x_b}^{x_e}\Bigl(\omega^2\hat {\cal G}_x(x_A,x,\omega)\Delta\alpha(x)\hat P(x,\omega)+
\frac{i\omega}{\beta_A(x)}\{\partial_x\hat {\cal G}_x(x_A,x,\omega)\}\Delta\beta(x)\hat Q(x,\omega)\Bigr){\rm d}x\nonumber\\
&&\hspace{-.7cm}+\Bigl(i\omega\hat {\cal G}_x(x_A,x,\omega)\hat Q(x,\omega)-\frac{1}{\beta_A(x)}\{\partial_x\hat {\cal G}_x(x_A,x,\omega)\}\hat P(x,\omega)\Bigr)\Bigr|_{x_b}^{x_e},\label{eq61k}
\end{eqnarray}
with
\begin{eqnarray}
\Delta\alpha(x)&=&\alpha(x)-\alpha_A(x),\\
\Delta\beta(x)&=&\beta(x)-\beta_A(x),
\end{eqnarray}
and where $\chi(x_A)$ is the  characteristic function, defined as
\begin{eqnarray}
\chi(x_A)=\begin{cases}
&1\quad\mbox{for}\quad x_b<x_A<x_e,\\
&\frac12 \quad\mbox{for}\quad x_A=x_b \quad\mbox{or}\quad x_A=x_e,\\
&0 \quad\mbox{for}\quad x_A<x_b \quad\mbox{or}\quad x_A>x_e.
\end{cases}
\end{eqnarray}
Equation (\ref{eq61k}) is a generalisation of equation (\ref{eq31k}), transformed to the frequency domain.
It expresses the wave field at any point $x_A$ between $x_b$ and $x_e$ (including these points). 
The first term on the right-hand side accounts for the contribution of the sources between
$x_b$ and $x_e$, the second term describes scattering due to the material contrast functions $\Delta\alpha(x)$ and $\Delta\beta(x)$, 
and the last term describes contributions from the fields at $x_b$ and $x_e$. 
The  representation of equation (\ref{eq61k}) finds applications in the analysis of wave scattering problems in space-dependent materials 
\citep{Hoop95Book, Morse53Book, Bleistein84Book, Born65Book, Oristaglio89IP}.

\subsection{Time-dependent material}\label{sec7.2}

We derive a general wave field representation for a time-dependent material  with piecewise continuous parameters $\alpha(t)$ and $\beta(t)$.
Our starting point is the global reciprocity theorem of the space-convolution type for a time-dependent material, formulated by equation (\ref{eq1017t}).
In anticipation of using the reciprocity relation (\ref{eq1022ta}), for  state $A$ we take the acausal Green's state, hence, we substitute 
$-ik\check b_A(k,t)=\delta(t-t_A)$, $\check a_A(k,t)=0$,
$\check U_A(k,t)=\check {\cal G}_t^a(k,t,t_A)$ and $\check V_A(k,t)=-\frac{\beta_A(t)}{ik}\partial_t\check {\cal G}_t^a(k,t,t_A)$.
For state $B$ we take the actual field and drop the subscripts $B$. 
Substitution into equation (\ref{eq1017t}), using reciprocity relation (\ref{eq1022ta}), this yields the following representation
\begin{eqnarray}
&&\hspace{-.7cm}\chi(t_A)\check U(k,t_A)=\int_{t_b}^{t_e}\Bigl(-ik\check {\cal G}_t(k,t_A,t)\check b(k,t)-
\beta_A(t)\{\partial_t\check {\cal G}_t(k,t_A,t)\}\check a(k,t)\Bigr){\rm d}t
\nonumber\\
&&\hspace{-.7cm}+\int_{t_b}^{t_e}\Bigl(-k^2\check {\cal G}_t(k,t_A,t)\Delta\alpha^{-1}(t)\check U(k,t)+
ik\beta_A(t)\{\partial_t\check {\cal G}_t(k,t_A,t)\}\Delta\beta^{-1}(t)\check V(k,t)\Bigr){\rm d}t\nonumber\\
&&\hspace{-.7cm}+\Bigl(ik\check {\cal G}_t(k,t_A,t)\check V(k,t)+\beta_A(t)\{\partial_t\check {\cal G}_t(k,t_A,t)\}\check U(k,t)\Bigr)\Bigr|_{t_b}^{t_e},\label{eq65k}
\end{eqnarray}
with
\begin{eqnarray}
\Delta\alpha^{-1}(t)&=&\alpha^{-1}(t)-\alpha_A^{-1}(t),\\
\Delta\beta^{-1}(t)&=&\beta^{-1}(t)-\beta_A^{-1}(t),
\end{eqnarray}
and where $\chi(t_A)$ is the  characteristic function, defined as
\begin{eqnarray}
\chi(t_A)=\begin{cases}
&1\quad\mbox{for}\quad t_b<t_A<t_e,\\
&\frac12 \quad\mbox{for}\quad t_A=t_b \quad\mbox{or}\quad t_A=t_e,\\
&0 \quad\mbox{for}\quad t_A<t_b \quad\mbox{or}\quad t_A>t_e.
\end{cases}
\end{eqnarray}
Equation (\ref{eq65k}) is a generalisation of equation (\ref{eq38k}), transformed to the wavenumber domain.
It expresses the wave field at any time $t_A$ between $t_b$ and $t_e$ (including these time instants). 
The first term on the right-hand side accounts for the contribution of the sources between
$t_b$ and $t_e$, the second term describes scattering due to the material contrast functions $\Delta\alpha^{-1}(t)$ and $\Delta\beta^{-1}(t)$, 
and the last term describes contributions from the fields at $t_b$ and $t_e$
(with the contribution from the field at $t_e$ being zero when $t_A<t_e$). 
The representation of equation (\ref{eq65k}) finds potential applications in the analysis of wave scattering problems in time-dependent materials.

\section{Green's function retrieval}

\subsection{Space-dependent material}\label{sec8.1}

Under specific circumstances, the correlation of passive wave measurements at two receivers yields the response to a virtual impulsive source at the position
of one of these receivers, observed by the other receiver (i.e., the Green's function between the receivers).
This concept has found numerous applications in
ultrasonics \citep{Weaver2001PRL, Weaver2002Ultrasonics, Malcolm2004PRE},
seismology \citep{Campillo2003Science, Wapenaar2003GEO, Snieder2004PRE, Schuster2004GJI, VanManen2005PRL, Shapiro2005Science, Lin2009GJI, Ruigrok2012GRL},
ocean acoustics \citep{Roux2004JASA2, Sabra2005JASA},
infrasound \citep{Haney2009GRL, Fricke2014JGR},
medical imaging \citep{Sabra2007APL, Benech2013JASA}
and engineering \citep{Snieder2006BSSA, Kohler2007BSSA}. 

Following the approach of  \cite{Wapenaar2006GEO}, 
we use the global reciprocity theorem of the time-correlation type (equation (\ref{eq1017c})) to derive an expression for Green's function retrieval
for a space-dependent material  with piecewise continuous parameters $\alpha(x)$ and $\beta(x)$. 
To this end, we take identical material parameters in both states, i.e., $\alpha_A=\alpha_B=\alpha$ and  $\beta_A=\beta_B=\beta$ and
we assume that the material is homogeneous for $x\le x_b$ and for $x\ge x_e$. 
For state $A$ we take a Green's state with a  unit source at $x_A$ between $x_b$ and $x_e$, and we substitute  
 $-i\omega\hat a_A(x,\omega)=\delta(x-x_A)$, $\hat b_A(x,\omega)=0$ and $\hat P_A(x,\omega)=\hat {\cal G}_x(x,x_A,\omega)$;
for $\hat Q_A(x,\omega)$ we use equations (\ref{eq54a}) and (\ref{eq55a}).
For state $B$ we take  a Green's state with a unit source at $x_B$ between $x_b$ and $x_e$, and we substitute similar expressions.
Furthermore, we use the source-receiver reciprocity relation of
equation (\ref{eq1022}). This yields
\begin{eqnarray}
&&\hspace{-.8cm}2i\Im\{\hat {\cal G}_x(x_B,x_A,\omega)\}= \nonumber\\
&&\hspace{.4cm} \frac{2i\omega}{\eta(x_b)}\hat {\cal G}_x^*(x_A,x_b,\omega)\hat {\cal G}_x(x_B,x_b,\omega)+ \frac{2i\omega}{\eta(x_e)}\hat {\cal G}_x^*(x_A,x_e,\omega)\hat {\cal G}_x(x_B,x_e,\omega),\label{eq1025}
\end{eqnarray}
where $\Im$ denotes the imaginary part.
Applying an inverse temporal Fourier transform yields
\begin{eqnarray}
&&\hspace{-.8cm}{\cal G}_x(x_B,x_A,t)-{\cal G}_x(x_B,x_A,-t)=\nonumber \\
&&\hspace{.4cm} 
-2\partial_t\sum_{x=x_b,x_e}\frac{1}{\eta(x)}\int_{-\infty}^\infty {\cal G}_x(x_A,x,t'){\cal G}_x(x_B,x,t+t') {\rm d}t'.\label{eq1025b}
\end{eqnarray}
The right-hand side is  the time derivative of the superposition of  time correlations of 
measurements by receivers at positions $x_A$ and $x_B$, in response to  impulsive sources at $x_b$ and $x_e$.
The left-hand side is the causal Green's function ${\cal G}_x(x_B,x_A,t)$ between $x_A$ and $x_B$, minus its time-reversed version.
Hence, the Green's function  ${\cal G}_x(x_B,x_A,t)$ is retrieved by evaluating the right-hand side of this expression and taking the causal part.
Note that this is independent of the positions $x_b$ and $x_e$ of the sources, as long as the receivers are located
between these sources and the material left of $x_b$ and right of $x_e$ is homogeneous. When the impulsive sources at $x_b$ and $x_e$
are replaced by uncorrelated noise sources, the retrieved response is the Green's function ${\cal G}_x(x_B,x_A,t)$,
convolved with the autocorrelation of the noise.

\subsection{Time-dependent material}\label{sec8.2}

Following a similar approach as in section \ref{sec8.1}, we use the global reciprocity theorem of the space-correlation type (equation (\ref{eq1017tc})) 
to derive an expression for Green's function retrieval for a time-dependent material  with piecewise continuous parameters $\alpha(t)$ and $\beta(t)$. 
We take again $\alpha_A=\alpha_B=\alpha$ and  $\beta_A=\beta_B=\beta$.
For state $A$ we take  an acausal Green's state with a unit sink at $t_A$ 
between $t_b$ and $t_e$ and we substitute $-ik\check b_A(k,t)=\delta(t-t_A)$, $\check a_A(k,t)=0$,
$\check U_A(k,t)=\check {\cal G}_t^a(k,t,t_A)$ and $\check V_A(k,t)=-\frac{\beta(t)}{ik}\partial_t\check {\cal G}_t^a(k,t,t_A)$. 
For state $B$ we take an acausal Green's state with a unit sink at $t_B$ between $t_b$ and $t_e$, and we substitute similar expressions.
Furthermore, we use the source-receiver reciprocity relation of equation (\ref{eq1022ta}). This yields
\begin{eqnarray}
&&\hspace{-.8cm}\check {\cal G}_t(k,t_B,t_A)-\{\check {\cal G}_t^a(k,t_B,t_A)\}^*=\nonumber\\
&&\hspace{0.4cm}\beta(t_b)\Bigl[\check {\cal G}_t^*(k,t_A,t_b)\partial_{t_b}\check {\cal G}_t(k,t_B,t_b) -\{\partial_{t_b}\check {\cal G}_t^*(k,t_A,t_b)\}\check {\cal G}_t(k,t_B,t_b) \Bigr].\label{eq1052}
\end{eqnarray}
Applying an inverse spatial Fourier transform yields
\begin{eqnarray}
&&\hspace{-.8cm}{\cal G}_t(x,t_B,t_A)-{\cal G}_t^a(-x,t_B,t_A)=\label{eq1052b}\\
&&\hspace{0.4cm}\beta(t_b)\int_{-\infty}^\infty\Bigl[{\cal G}_t(x',t_A,t_b)\partial_{t_b} {\cal G}_t(x+x',t_B,t_b) - \{\partial_{t_b}{\cal G}_t(x',t_A,t_b)\}{\cal G}_t(x+x',t_B,t_b)\Bigr]{\rm d}x'.\nonumber
\end{eqnarray}
The right-hand side is   the superposition of  space correlations of 
measurements by receivers at time instants $t_A$ and $t_B$, in response to an impulsive source at $t_b$ and its time derivative and vice-versa.
The left-hand side is the causal Green's function ${\cal G}_t(x,t_B,t_A)$ between $t_A$ and $t_B$, minus its space-reversed acausal counterpart.
Hence,  the Green's function ${\cal G}_t(x,t_B,t_A)$ (when $t_B>t_A$) or $-{\cal G}_t^a(-x,t_B,t_A)$ (when $t_B<t_A$) 
 is retrieved by evaluating the right-hand side of this expression.
Note that this is independent of the time instant $t_b$ of the source, as long $t_A$ and $t_B$ are both larger than $t_b$.
Unlike the two-sided representation of equation (\ref{eq1025b}), which requires sources at $x_b$ and $x_e$, this is a single-sided representation, which 
requires sources at $t_b$ only.
Note that the time-derivatives in equation (\ref{eq1052b}) act on a superposition of left- and right-going waves at $t_b$
(see for example Figure \ref{Fig2}, with $t_b=t_0$), hence, we cannot use an expression similar to equation (\ref{eq54a}) to 
 simplify the right-hand side of equation (\ref{eq1052b}) further.

\section{Matrix-vector wave equation}

\subsection{Space-dependent material}

For a space-dependent material with continuous parameters $\alpha(x)$ and $\beta(x)$,
equations (\ref{eq6}) and (\ref{eq7}) can be combined into the following matrix-vector wave equation in the $x,t$-domain 
\citep{Corones75JMAA, Kosloff83GEO, Fishman84JMP, Ursin83GEO, Wapenaar86GP2, Loseth2007GJI}
\begin{eqnarray}
\partial_x{\bf q}_x={\bf A}_x{\bf q}_x+{\bf d}_x,\label{eq8}
\end{eqnarray}
with wave field vector ${\bf q}_x(x,t)$, operator matrix ${\bf A}_x(x,t)$ and source vector ${\bf d}_x(x,t)$ defined as
\begin{eqnarray}
{\bf q}_x=\begin{pmatrix}  P\\  Q\end{pmatrix},\quad
{\bf A}_x=\begin{pmatrix} 0 & -\beta\partial_t \\ -\alpha\partial_t & 0 \end{pmatrix},\quad
{\bf d}_x=\begin{pmatrix}  b\\  a\end{pmatrix}.\label{eq9}
\end{eqnarray}
For a space-dependent material with piecewise continuous parameters, this equation is supplemented with boundary conditions at all points where $\alpha(x)$ and $\beta(x)$ 
are discontinuous. The boundary condition is that ${\bf q}_x(x,t)$ is continuous at those points.

Using the temporal Fourier transform defined in equation (\ref{eq10}), we obtain the following matrix-vector wave equation in the $x,\omega$-domain
\begin{eqnarray}
\partial_x\hat{\bf q}_x=\hat{\bf A}_x\hat{\bf q}_x+\hat{\bf d}_x,\label{eq17}
\end{eqnarray}
with wave field vector $\hat{\bf q}_x(x,\omega)$, matrix $\hat{\bf A}_x(x,\omega)$ and source vector $\hat{\bf d}_x(x,\omega)$ defined as
\begin{eqnarray}
\hat{\bf q}_x=\begin{pmatrix}  \hat P\\  \hat Q\end{pmatrix},\quad
\hat{\bf A}_x=\begin{pmatrix} 0 &  i\omega\beta\\ i\omega\alpha& 0 \end{pmatrix},\quad
\hat{\bf d}_x=\begin{pmatrix}  \hat b\\  \hat a\end{pmatrix}.\label{eq18}
\end{eqnarray}
Note that matrix $\hat{\bf A}_x(x,\omega)$ obeys the following symmetry properties
\begin{eqnarray}
\hat{\bf A}_x^t{\bf N}&=&-{\bf N}\hat{\bf A}_x,\label{eq19a}\\
\hat{\bf A}_x^\dagger{\bf K}&=&-{\bf K}\hat{\bf A}_x,\label{eq19b}\\
\hat{\bf A}_x^*{\bf J}&=&{\bf J}\hat{\bf A}_x,\label{eq19c}
\end{eqnarray}
where superscript $t$ denotes transposition, superscript $\dagger$ denotes transposition and complex conjugation, and where
\begin{eqnarray}\label{eq4.3}
{{\bf N}}=\begin{pmatrix} 0 & 1 \\ -1 & 0 \end{pmatrix},
\quad {{\bf K}}=\begin{pmatrix} 0 & 1 \\ 1 & 0 \end{pmatrix},
\quad {{\bf J}}=\begin{pmatrix} 1 & 0 \\ 0 & -1 \end{pmatrix}.\label{eq21a}
\end{eqnarray}

\subsection{Time-dependent material}

Applying the mapping of equation (\ref{eqmap}) to equations (\ref{eq8}) and (\ref{eq9}) yields the following matrix-vector wave equation in the $x,t$-domain
for a time-dependent material with continuous parameters $\alpha(t)$ and $\beta(t)$
\begin{eqnarray}
\partial_t{\bf q}_t={\bf A}_t{\bf q}_t+{\bf d}_t,\label{eq58}
\end{eqnarray}
with wave field vector ${\bf q}_t(x,t)$, operator matrix ${\bf A}_t(x,t)$ and source vector ${\bf d}_t(x,t)$ defined as
\begin{eqnarray}
{\bf q}_t=\begin{pmatrix}  U\\  V\end{pmatrix},\quad
{\bf A}_t=\begin{pmatrix} 0 & -\frac{1}{\beta}\partial_x \\ -\frac{1}{\alpha}\partial_x & 0 \end{pmatrix},\quad
{\bf d}_t=\begin{pmatrix}  a\\  b\end{pmatrix}.\label{eq59}
\end{eqnarray}
For a time-dependent material with piecewise continuous parameters, this equation is supplemented with boundary conditions at all time instants 
where $\alpha(t)$ and $\beta(t)$ 
are discontinuous. The boundary condition is that ${\bf q}_t(x,t)$ is continuous at those time instants.

Applying the mapping of equation (\ref{eqmap2}) to equations (\ref{eq17}) and (\ref{eq18}), we obtain the following matrix-vector wave equation in the $k,t$-domain
\begin{eqnarray}
\partial_t\check{\bf q}_t=\check{\bf A}_t\check{\bf q}_t+\check{\bf d}_t,\label{eq67}
\end{eqnarray}
with
\begin{eqnarray}
\check{\bf q}_t=\begin{pmatrix}  \check U\\  \check V\end{pmatrix},\quad
\check{\bf A}_t=\begin{pmatrix} 0 & -\frac{ik}{\beta}\\-\frac{ik}{\alpha}& 0 \end{pmatrix},\quad
\check{\bf d}_t=\begin{pmatrix}  \check a\\  \check b\end{pmatrix}.\label{eq68}
\end{eqnarray}
Matrix $\check{\bf A}_t$ obeys the same symmetries as $\hat{\bf A}_x$, as formulated by equations (\ref{eq19a})--(\ref{eq19c}).

\section{Propagator matrices and representations}\label{sec10}

\subsection{Space-dependent material}\label{sec10.1}

For a space-dependent material, a propagator matrix ``propagates'' a wave field (represented as a vectorial quantity) 
from one plane in space to another  \citep{Thomson50JAP, Haskell53BSSA, Gilbert66GEO}.
It has found many applications, particularly in elastodynamic wave problems \citep{Kennett83Book, Kennett72GJRAS, Woodhouse74GJR}.

We define the propagator matrix ${\bf W}_x(x,x_0,t)$ for a space-dependent material with continuous parameters $\alpha(x)$ and $\beta(x)$
as the solution of matrix-vector equation (\ref{eq8}) without the source term, hence
\begin{eqnarray}
\partial_x{\bf W}_x={\bf A}_x{\bf W}_x,\label{eq121}
\end{eqnarray}
with operator matrix ${\bf A}_x(x,t)$ defined in equation (\ref{eq9}) and with boundary condition 
\begin{eqnarray}
{\bf W}_x(x_0,x_0,t)={\bf I}\delta(t),\label{eq114b}
\end{eqnarray}
where ${\bf I}$ is the identity matrix.

A simple representation for the wave field vector ${\bf q}_x(x,t)$ obeying equation (\ref{eq8})
 is obtained when ${\bf q}_x$ and ${\bf W}_x$ are defined in the same source-free 
material between $x_0$ and $x$ (where $x$ can be either larger or smaller than $x_0$). 
Whereas ${\bf q}_x(x,t)$ can have any time-dependency at $x=x_0$,  
${\bf W}_x(x,x_0,t)$ collapses to ${\bf I}\delta(t)$ at $x=x_0$. Because equations (\ref{eq8}) and (\ref{eq121}) are linear,
a representation for ${\bf q}_x(x,t)$ follows by applying Huygens' superposition principle, according to 
\begin{eqnarray}
{\bf q}_x(x,t)= \int_{-\infty}^\infty {\bf W}_x(x,x_0,t-t'){\bf q}_x(x_0,t'){\rm d}t'.\label{eq90t}
\end{eqnarray}
Note that ${\bf W}_x(x,x_0,t)$ propagates the wave field vector ${\bf q}_x$ from $x_0$ to $x$, hence the name ``propagator matrix''. 
We partition ${\bf W}_x(x,x_0,t)$ as follows
\begin{eqnarray}
{\bf W}_x(x,x_0,t)=
\begin{pmatrix}W_x^{P,P} & W_x^{P,Q}\\W_x^{Q,P} & W_x^{Q,Q}\end{pmatrix}(x,x_0,t).\label{eq90bc}
\end{eqnarray}
The first and second superscripts refer to the wave field quantities in vector ${\bf q}_x$, defined in equation (\ref{eq9}), at $x$ and $x_0$, respectively.
For more general representations with propagator matrices, including source terms and differences in material parameters
(analogous to the representation with Green's functions discussed in section \ref{sec7.1}), see  \cite{Wapenaar2022JASA}.

\begin{figure}
\centerline{\epsfysize=7. cm \epsfbox{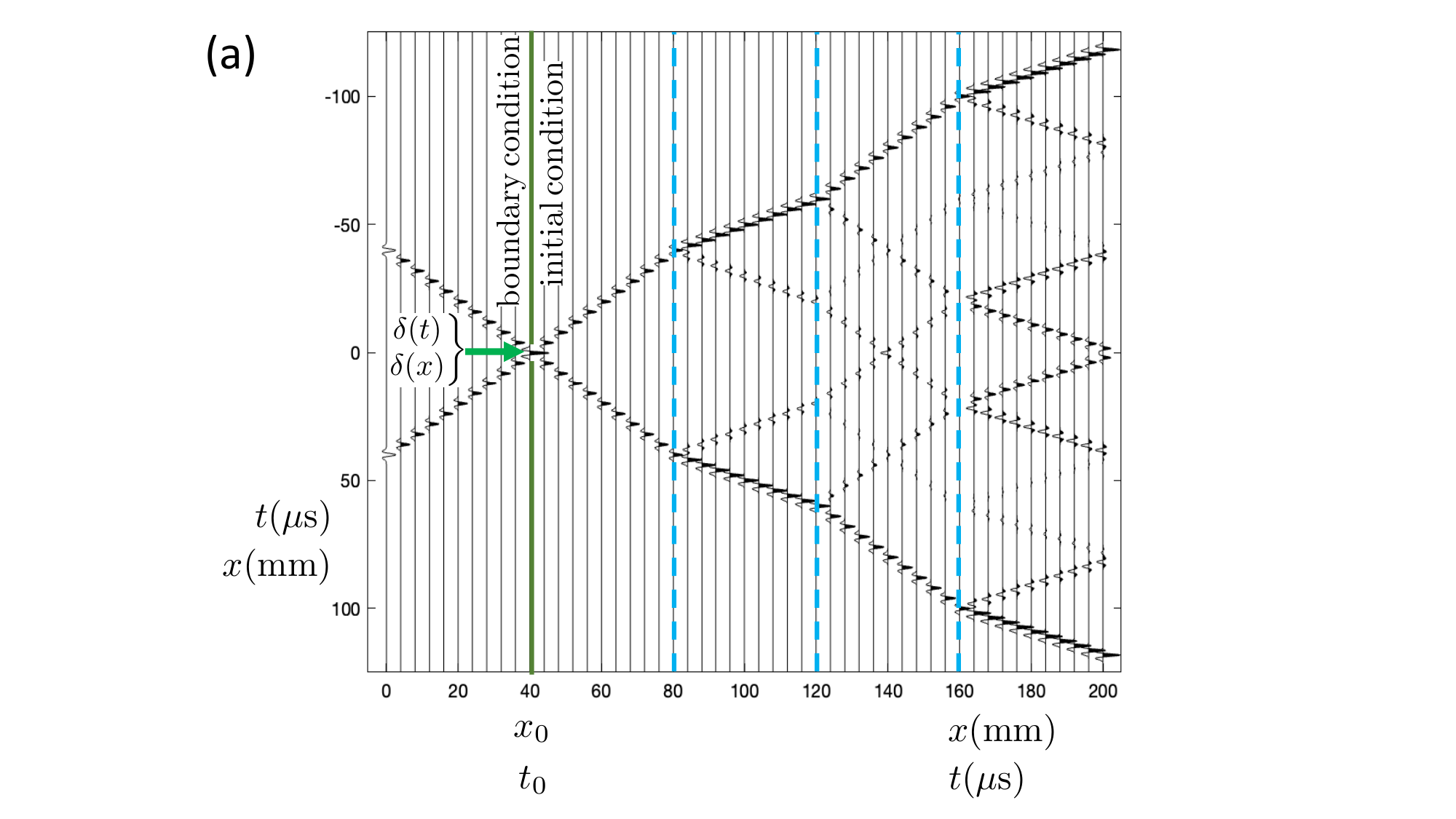}}
\centerline{\epsfysize=7. cm \epsfbox{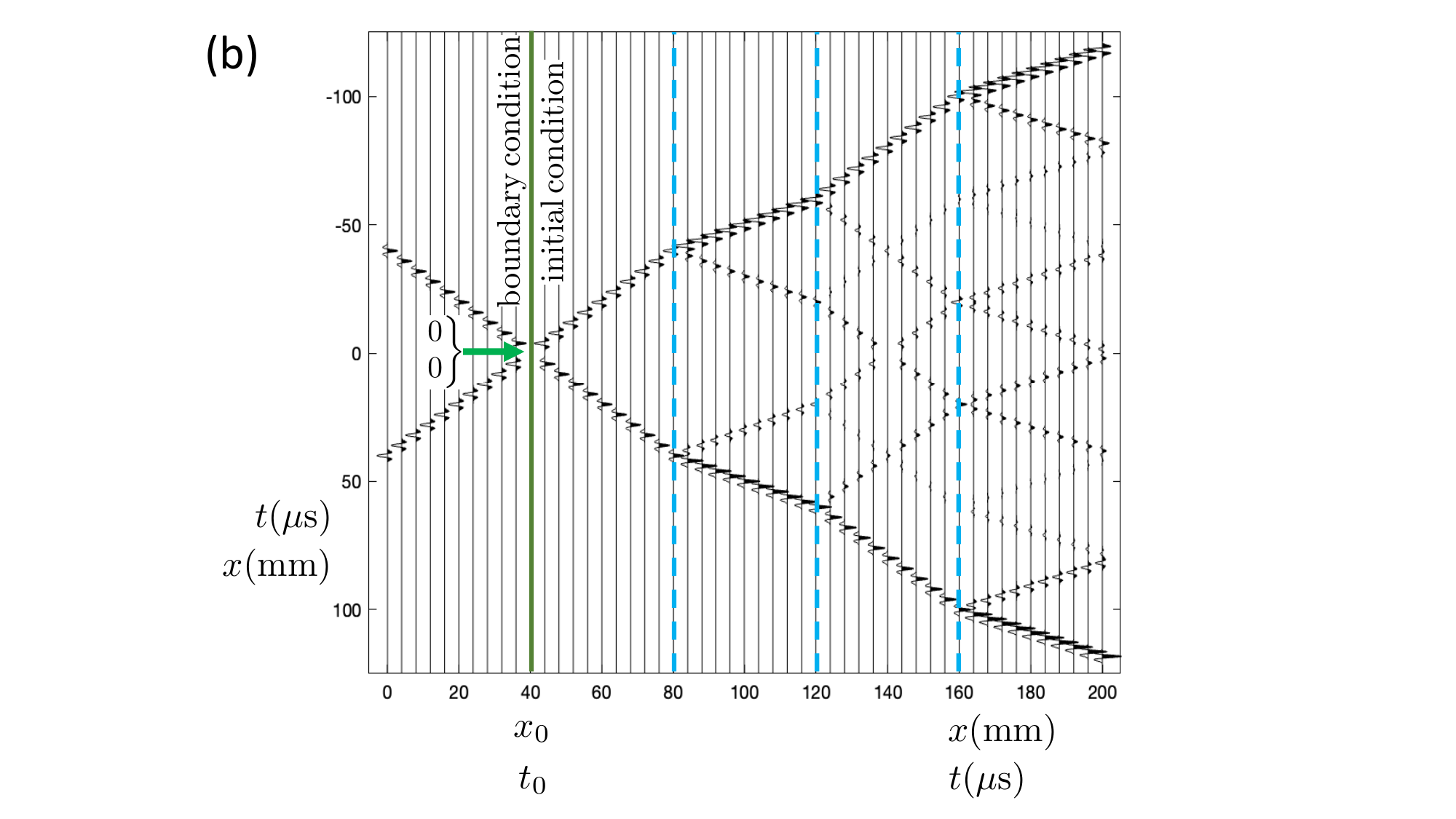}}
\caption{Propagator matrix elements $W_x^{P,P}(x,x_0,t)$ (a) and $W_x^{P,Q}(x,x_0,t)$ (b) 
(convolved with a temporal  wavelet) for a piecewise homogeneous space-dependent material.
The labels at the vertical axes denote time (in $\mu$s) and those at the horizontal axes denote space (in mm).
With interchanged labels (and ``boundary condition'' replaced by ``initial condition'') these figures can be interpreted as $W_t^{U,U}(x,t,t_0)$ (a) and $W_t^{U,V}(x,t,t_0)$ (b) 
(convolved with a spatial  wavelet) for a piecewise constant time-dependent material.}\label{Fig3}
\end{figure}

Using the temporal Fourier transform defined in equation (\ref{eq10}), we obtain the following
space-frequency domain equation for $\hat{\bf W}_x(x,x_0,\omega)$
\begin{eqnarray}
\partial_x\hat{\bf W}_x=\hat{\bf A}_x\hat{\bf W}_x,\label{eq124}
\end{eqnarray}
with matrix $\hat{\bf A}_x(x,\omega)$ defined in equation (\ref{eq18}) and with boundary condition 
\begin{eqnarray}
\hat{\bf W}_x(x_0,x_0,\omega)={\bf I}.\label{eq125}
\end{eqnarray}
The representation of equation (\ref{eq90t}) transforms to
\begin{eqnarray}
\hat{\bf q}_x(x,\omega)=\hat{\bf W}_x(x,x_0,\omega)\hat{\bf q}_x(x_0,\omega).\label{eq127}
\end{eqnarray}
By applying this equation recursively, it follows that $\hat{\bf W}_x$ obeys the following recursive expression
\begin{eqnarray}
\hat{\bf W}_x(x_N,x_0,\omega)=\hat{\bf W}_x(x_N,x_{N-1},\omega)\cdots\hat{\bf W}_x(x_n,x_{n-1},\omega)\cdots\hat{\bf W}_x(x_1,x_0,\omega),
\label{eq94}
\end{eqnarray}
where $x_1\cdots x_n\cdots x_{N-1}$ are points where the material parameters may be discontinuous.
As a special case of equation (\ref{eq94}) we obtain
\begin{eqnarray}
\hat{\bf W}_x(x_{n-1},x_n,\omega)\hat{\bf W}_x(x_n,x_{n-1},\omega)=\hat{\bf W}_x(x_{n-1},x_{n-1},\omega)={\bf I},\label{eq95}
\end{eqnarray}
from which it follows that $\hat{\bf W}_x(x_{n-1},x_n,\omega)$ is the inverse of $\hat{\bf W}_x(x_n,x_{n-1},\omega)$.
For a homogeneous slab between $x_{n-1}$ and $x_n$, with parameters $\alpha_n$, $\beta_n$, $c_n=1/\sqrt{\alpha_n\beta_n}$, $\eta_n=\sqrt{\beta_n/\alpha_n}$
and thickness $\Delta x_n=x_n-x_{n-1}$, we have
\begin{eqnarray}
\hat W_x^{P,P}(x_n,x_{n-1},\omega)&=&\cos(\omega \Delta x_n/c_n),\label{eq96a}\\
\hat W_x^{P,Q}(x_n,x_{n-1},\omega)&=&i\eta_n\sin(\omega \Delta x_n/c_n),\\
\hat W_x^{Q,P}(x_n,x_{n-1},\omega)&=&\frac{i}{\eta_n}\sin(\omega \Delta x_n/c_n),\\
\hat W_x^{Q,Q}(x_n,x_{n-1},\omega)&=&\cos(\omega \Delta x_n/c_n).\label{eq99a}
\end{eqnarray}
From equation (\ref{eq94}), we obtain a similar recursive expression in the space-time domain, according to
\begin{eqnarray}
&&\hspace{-.7cm}{\bf W}_x(x_N,x_0,t)={\bf W}_x(x_N,x_{N-1},t)*_t\cdots*_t{\bf W}_x(x_n,x_{n-1},t)*_t\cdots*_t{\bf W}_x(x_1,x_0,t),
\label{eq98}
\end{eqnarray}
where $*_t$ denotes a time convolution (more formally defined in equation (\ref{eq90t})). 
For a homogeneous slab between $x_{n-1}$ and $x_n$, we find from equations (\ref{eq96a})--(\ref{eq99a})
\begin{eqnarray}
W_x^{P,P}(x_n,x_{n-1},t)&=&\frac{1}{2}\{\delta(t-\Delta x_n/c_n)+\delta(t+\Delta x_n/c_n)\},\label{eq99}\\
W_x^{P,Q}(x_n,x_{n-1},t)&=&\frac{\eta_n}{2}\{\delta(t-\Delta x_n/c_n)-\delta(t+\Delta x_n/c_n)\},\\
W_x^{Q,P}(x_n,x_{n-1},t)&=&\frac{1}{2\eta_n}\{\delta(t-\Delta x_n/c_n)-\delta(t+\Delta x_n/c_n)\},\\
W_x^{Q,Q}(x_n,x_{n-1},t)&=&\frac{1}{2}\{\delta(t-\Delta x_n/c_n)+\delta(t+\Delta x_n/c_n)\}.\label{eq99d}
\end{eqnarray}
For the same piecewise homogeneous material as used for the numerical example in section \ref{sec4.1}, 
the elements $W_x^{P,P}(x,x_0,t)$ and $W_x^{P,Q}(x,x_0,t)$ for $x_0=40$ mm 
(convolved with a temporal wavelet with a central frequency $\omega_0/2\pi=300$ kHz)
are shown as $x,t$-diagrams in Figures \ref{Fig3}a and \ref{Fig3}b. The green lines indicate the boundary conditions 
$W_x^{P,P}(x_0,x_0,t)=\delta(t)$ and $W_x^{P,Q}(x_0,x_0,t)=0$ (equations (\ref{eq114b}) and (\ref{eq90bc})). 
Note that these figures clearly exhibit the recursive character, described by equation (\ref{eq98}).

\subsection{Time-dependent material}\label{sec10.2}

For a time-dependent material, a propagator matrix propagates a wave field from one instant in time to another.
In the literature on time-dependent materials this matrix is usually called the transfer matrix \citep{Torrent2018PRB, Salem2015arXiv, Pacheco2020NP},
but for consistency with section \ref{sec10.1}, we hold on to the name propagator matrix. 

We define the propagator matrix ${\bf W}_t(x,t,t_0)$ for a time-dependent material 
with continuous parameters $\alpha(t)$ and $\beta(t)$ as the solution of matrix-vector equation (\ref{eq58})
without the source term, hence
\begin{eqnarray}
\partial_t{\bf W}_t={\bf A}_t{\bf W}_t,\label{eq171}
\end{eqnarray}
with operator matrix ${\bf A}_t(x,t)$ defined in equation (\ref{eq59}) and with initial condition 
\begin{eqnarray}
{\bf W}_t(x,t_0,t_0)={\bf I}\delta(x).\label{eq136b}
\end{eqnarray}
Note that the mapping of equation (\ref{eqmap}) not only applies to the wave equations (equations (\ref{eq121}) and (\ref{eq171})), but also to the boundary and initial conditions 
(equations (\ref{eq114b}) and (\ref{eq136b})). Consequently, the mappings of equations  (\ref{eqmap}) and  (\ref{eqmap2}) 
also apply to all expressions for the propagator matrix in the space-time and Fourier-domains, respectively.
We discuss a few of these mappings explicitly.

The representation of equation (\ref{eq90t}) maps to
\begin{eqnarray}
{\bf q}_t(x,t)= \int_{-\infty}^\infty {\bf W}_t(x-x',t,t_0){\bf q}_t(x',t_0){\rm d}x'.\label{eq90x}
\end{eqnarray}
Note that ${\bf W}_t(x,t,t_0)$ propagates the wave field vector ${\bf q}_t$ from $t_0$ to $t$.
We partition ${\bf W}_t(x,t,t_0)$ as follows
\begin{eqnarray}
{\bf W}_t(x,t,t_0)=
\begin{pmatrix}W_t^{U,U} & W_t^{U,V}\\W_t^{V,U} & W_t^{V,V}\end{pmatrix}(x,t,t_0).\label{eq108}
\end{eqnarray}
The first and second superscripts refer to the wave field quantities in vector ${\bf q}_t$, defined in equation (\ref{eq59}), at $t$ and $t_0$, respectively.
The recursive expression of equation (\ref{eq98}) maps to
\begin{eqnarray}
&&\hspace{-.7cm}{\bf W}_t(x,t_N,t_0)={\bf W}_t(x,t_N,t_{N-1})*_x\cdots*_x{\bf W}_t(x,t_n,t_{n-1})*_x\cdots*_x{\bf W}_t(x,t_1,t_0),
\label{eq98x}
\end{eqnarray}
where $*_x$ denotes a space convolution (more formally defined in equation (\ref{eq90x})). 
For a constant slab between $t_{n-1}$ and $t_n$ with duration $\Delta t_n=t_n-t_{n-1}$, we find from equations (\ref{eq99}) -- (\ref{eq99d})
\begin{eqnarray}
W_t^{U,U}(x,t_n,t_{n-1})&=&\frac{1}{2}\{\delta(x-c_n\Delta t_n)+\delta(x+c_n\Delta t_n)\},\\
W_t^{U,V}(x,t_n,t_{n-1})&=&\frac{1}{2\eta_n}\{\delta(x-c_n\Delta t_n)-\delta(x+c_n\Delta t_n)\},\\
W_t^{V,U}(x,t_n,t_{n-1})&=&\frac{\eta_n}{2}\{\delta(x-c_n\Delta t_n)-\delta(x+c_n\Delta t_n)\},\\
W_t^{V,V}(x,t_n,t_{n-1})&=&\frac{1}{2}\{\delta(x-c_n\Delta t_n)+\delta(x+c_n\Delta t_n)\}.
\end{eqnarray}
For the same piecewise constant material as used for the first numerical example in section \ref{sec4.2},
the elements $W_t^{U,U}(x,t,t_0)$ and $W_t^{U,V}(x,t,t_0)$ for $t_0=40$ $\mu$s (convolved with a spatial wavelet with a central wavenumber $k_0/2\pi=300*10^3$ km$^{-1}$)
are shown as $x,t$-diagrams in Figures \ref{Fig3}a and \ref{Fig3}b.
The green lines indicate the initial conditions  $W_t^{U,U}(x,t_0,t_0)=\delta(x)$ and $W_t^{U,V}(x,t_0,t_0)=0$ (equations (\ref{eq136b}) and (\ref{eq108})).
Note that these figures clearly exhibit the recursive character, described by equation (\ref{eq98x}).

Finally we show that the Green's function ${\cal G}_t(x,t,t_0)$, obeying equation (\ref{eq36b}) with causality condition (\ref{eqcaus2}),  is related to $W_t^{U,V}(x,t,t_0)$ via
\begin{eqnarray}
-\partial_x{\cal G}_t(x,t,t_0)=H(t-t_0)W_t^{U,V}(x,t,t_0).\label{eq93}
\end{eqnarray}
Due to the Heaviside function $H(t-t_0)$, the causality condition  (\ref{eqcaus2}) is fulfilled, 
so we only need to show that $H(t-t_0)W_t^{U,V}(x,t,t_0)$ obeys the same wave equation as 
$-\partial_x{\cal G}_t(x,t,t_0)$. For the first derivative with respect to time we obtain, using the product rule for differentiation and equations (\ref{eq59}), (\ref{eq171}), (\ref{eq136b}) and (\ref{eq108}),
\begin{eqnarray}
\partial_t\{H(t-t_0)W_t^{U,V}(x,t,t_0)\}=-\frac{1}{\beta(t)}H(t-t_0)\partial_xW_t^{V,V}(x,t,t_0).
\end{eqnarray}
Multiplying both sides with $\beta(t)$ and differentiating again with respect to time, we obtain in a similar way
\begin{eqnarray}
\partial_t\bigl(\beta(t)\partial_t\{H(t-t_0)W_t^{U,V}(x,t,t_0)\}\bigr)&=&-\partial_x\delta(x)\delta(t-t_0)\\
&+&\frac{1}{\alpha(t)}\partial_x^2\{H(t-t_0)W_t^{U,V}(x,t,t_0)\}.\nonumber
\end{eqnarray}
Comparing this with equation (\ref{eq36b}), with $c(t)$ 
defined in equation (\ref{eq8a}), we observe that $H(t-t_0)W_t^{U,V}(x,t,t_0)$ indeed obeys the same wave equation as $-\partial_x{\cal G}_t(x,t,t_0)$. Hence, 
${\cal G}_t(x,t,t_0)$  is obtained by integrating $-H(t-t_0)W_t^{U,V}(x,t,t_0)$  with respect to $x$ (see Figures  \ref{Fig2}a and  \ref{Fig3}b). 
Note that a relation similar to equation (\ref{eq93})  does not exist for a space-dependent material 
(since the causality conditions (equations (\ref{eqcaus1}) and (\ref{eqcaus2})) do not follow the mapping of equation (\ref{eqmap})). 

\section{Matrix-vector reciprocity theorems}

\subsection{Space-dependent material}\label{sec11.1}

We review matrix-vector reciprocity theorems for a space-dependent material with piecewise continuous parameters $\alpha(x)$ and $\beta(x)$.
We consider two independent states, indicated by subscripts $A$ and $B$, obeying equation (\ref{eq17}), and we derive relations between these states.
In the most general case, sources, material parameters and wave fields may be different in the two states. 
We consider the quantities $\partial_x\{\hat{\bf q}_{x,A}^t(x,\omega){\bf N}\hat{\bf q}_{x,B}(x,\omega)\}$ and 
$\partial_x\{\hat{\bf q}_{x,A}^\dagger(x,\omega){\bf K}\hat{\bf q}_{x,B}(x,\omega)\}$.
Applying the product rule for differentiation, using wave equation (\ref{eq17}) and symmetry relations (\ref{eq19a}) and (\ref{eq19b}) for states $A$ and $B$, yields
\begin{eqnarray}
\partial_x\{\hat {\bf q}_{x,A}^t{\bf N}\hat {\bf q}_{x,B}\}&=&
\hat {\bf q}_{x,A}^t{\bf N}\Delta\hat{\bf A}_x\hat {\bf q}_{x,B}+
\hat {\bf d}_{x,A}^t{\bf N}\hat {\bf q}_{x,B}+\hat {\bf q}_{x,A}^t{\bf N}\hat {\bf d}_{x,B},\label{eq112}
\label{eq201a}\\
\partial_x\{\hat {\bf q}_{x,A}^\dagger{\bf K}\hat {\bf q}_{x,B}\}&=&
\hat {\bf q}_{x,A}^\dagger{\bf K}\Delta\hat{\bf A}_x\hat {\bf q}_{x,B}+
\hat {\bf d}_{x,A}^\dagger{\bf K}\hat {\bf q}_{x,B}+\hat {\bf q}_{x,A}^\dagger{\bf K}\hat {\bf d}_{x,B},\label{eq113}
\label{eq221a}
\end{eqnarray}
with $\Delta\hat{\bf A}_x=\hat{\bf A}_{x,B}-\hat{\bf A}_{x,A}$. Equations (\ref{eq112}) and (\ref{eq113}) 
are the matrix-vector forms of the local reciprocity theorems of the time-convolution and
time-correlation type, respectively, as formulated by equations (\ref{eq1013}) and (\ref{eq1014}).
Integration of both sides of equations (\ref{eq112}) and (\ref{eq113}) from $x_b$ to $x_e$ yields 
\citep{Haines96JMP, Wapenaar96GJI1}
\begin{eqnarray}
&&\hspace{-.3cm}\hat {\bf q}_{x,A}^t{\bf N}\hat {\bf q}_{x,B}\Bigr|_{x_b}^{x_e}=
\int_{x_b}^{x_e}\{\hat {\bf q}_{x,A}^t{\bf N}\Delta\hat{\bf A}_x\hat {\bf q}_{x,B}+\hat {\bf d}_{x,A}^t{\bf N}\hat {\bf q}_{x,B}+\hat {\bf q}_{x,A}^t{\bf N}\hat {\bf d}_{x,B}\}{\rm d}x,
\label{eq202a}\\
&&\hspace{-.3cm}\hat {\bf q}_{x,A}^\dagger{\bf K}\hat {\bf q}_{x,B}\Bigr|_{x_b}^{x_e}=
\int_{x_b}^{x_e}\{\hat {\bf q}_{x,A}^\dagger{\bf K}\Delta\hat{\bf A}_x\hat {\bf q}_{x,B}+\hat {\bf d}_{x,A}^\dagger{\bf K}\hat {\bf q}_{x,B}+\hat {\bf q}_{x,A}^\dagger{\bf K}\hat {\bf d}_{x,B}\}{\rm d}x.
\label{eq222a}
\end{eqnarray}
These are the matrix-vector forms of the global reciprocity theorems of the time-convolution and
time-correlation type, respectively, as formulated by equations (\ref{eq1017}) and (\ref{eq1017c}).
When there are no sources and the material parameters are identical in both states, the right-hand sides of equations (\ref{eq112})--(\ref{eq222a}) are zero.
From equations (\ref{eq112}) and (\ref{eq113}) it then follows that 
$\hat {\bf q}_{x,A}^t{\bf N}\hat {\bf q}_{x,B}$ and $\hat {\bf q}_{x,A}^\dagger{\bf K}\hat {\bf q}_{x,B}$ are space-propagation invariants
\citep{Haines88GJI, Kennett90GJI, Koketsu91GJI, Takenaka93WM}.

We use equations (\ref{eq202a}) and (\ref{eq222a}) (with zeroes on the right-hand sides) to derive reciprocity relations for the propagator matrix. 
For state $A$ we substitute $\hat{\bf q}_{x,A}(x,\omega)=\hat{\bf W}_x(x,x_A,\omega)$ and, using equation (\ref{eq125}),
$\hat{\bf q}_{x,A}(x_A,\omega)={\bf I}$.
Similarly, For state $B$ we substitute $\hat{\bf q}_{x,B}(x,\omega)=\hat{\bf W}_x(x,x_B,\omega)$ and $\hat{\bf q}_{x,B}(x_B,\omega)={\bf I}$.
Taking $x_A$ and $x_B$ equal to $x_b$ and $x_e$ (in arbitrary order) yields
\begin{eqnarray}
\hat{\bf W}_x^t(x_B,x_A,\omega){\bf N}&=&{\bf N}\hat{\bf W}_x(x_A,x_B,\omega),\label{eq220}\\
\hat{\bf W}_x^\dagger(x_B,x_A,\omega){\bf K}&=&{\bf K}\hat{\bf W}_x(x_A,x_B,\omega).\label{eq220cor}\end{eqnarray}
Using ${\bf N}^{-1}{\bf K}=-{\bf J}$ we find from equations (\ref{eq220}) and (\ref{eq220cor})
\begin{eqnarray}
\hat{\bf W}_x^*(x_A,x_B,\omega){\bf J}={\bf J}\hat{\bf W}_x(x_A,x_B,\omega),\label{eq220corJ}
\end{eqnarray}
or, in the space-time domain
\begin{eqnarray}
{\bf W}_x(x_A,x_B,-t){\bf J}={\bf J}{\bf W}_x(x_A,x_B,t).\label{eq220corJst}
\end{eqnarray}

\subsection{Time-dependent material}\label{sec11.2}

The matrix-vector reciprocity theorems for a time-dependent material with piecewise continuous parameters $\alpha(t)$ and $\beta(t)$ follow from those in section \ref{sec11.1}
by applying the mappings of equations  (\ref{eqmap}) and  (\ref{eqmap2}). In particular, the reciprocity relations for the propagator matrix are
\begin{eqnarray}
\check{\bf W}_t^t(k,t_B,t_A){\bf N}&=&{\bf N}\check{\bf W}_t(k,t_A,t_B),\label{eq270}\\
\check{\bf W}_t^\dagger(k,t_B,t_A){\bf K}&=&{\bf K}\check{\bf W}_t(k,t_A,t_B),\label{eq270cor}\\
\check{\bf W}_t^*(k,t_A,t_B){\bf J}&=&{\bf J}\check{\bf W}_t(k,t_A,t_B),\label{eq270corJ}\\
{\bf W}_t(-x,t_A,t_B){\bf J}&=&{\bf J}{\bf W}_t(x,t_A,t_B).\label{eq270corJst}
\end{eqnarray}

\section{Marchenko-type focusing functions}

\subsection{Space-dependent material}\label{sec12.1}

\begin{figure}
\centerline{\epsfysize=7. cm \epsfbox{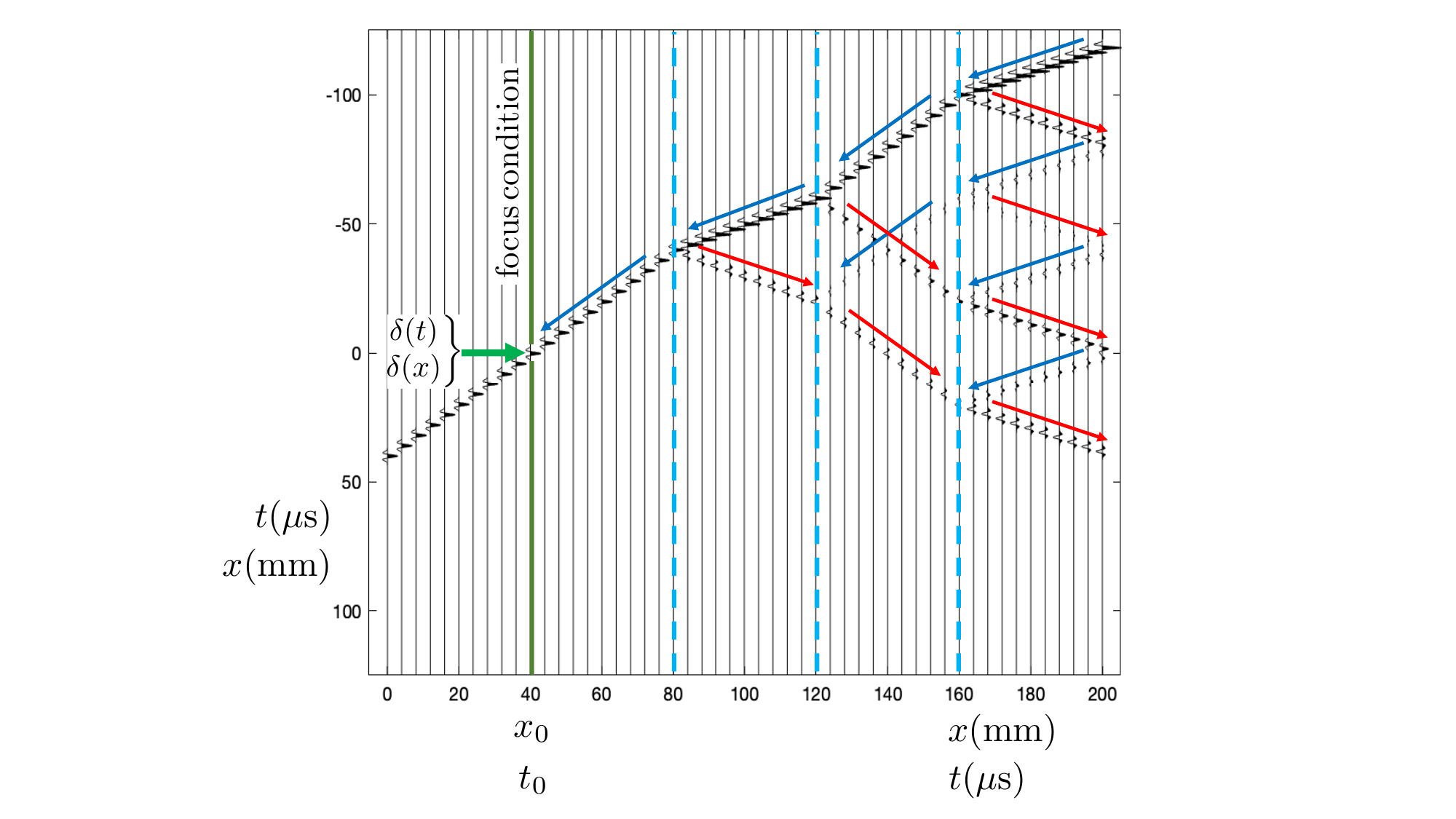}}
\caption{Focusing function $F_x^P(x,x_0,t)$ 
(convolved with a temporal wavelet) for a piecewise homogeneous space-dependent material.
The label at the vertical axis denotes time (in $\mu$s) and that at the horizontal axis denotes space (in mm).
With interchanged labels and reversed blue arrows this figure can be interpreted as $F_t^U(x,t,t_0)$ 
(convolved with a spatial wavelet) for a piecewise constant time-dependent material.}\label{Fig4}
\end{figure}

Building on work by \cite{Rose2001PRA, Rose2002IP}, geophysicists used the Marchenko equation to develop methods for retrieving 
the wave field inside a space-dependent material from reflection measurements at its boundary 
\citep{Broggini2012EJP, Broggini2012GEO, Wapenaar2014JASA, Ravasi2016GJI, Staring2018GEO, Jia2018GEO}.
Focusing functions play a central role in this methodology. 
For a 1D space-dependent material, a Marchenko-type focusing function $F_x(x,x_0,t)$ is defined as a specific solution of the wave equation, which focuses at the
focal point $x=x_0$  (i.e., $F_x(x_0,x_0,t)\propto \delta(t)$) and which propagates unidirectionally through the focal point.
It has recently been shown that there exists a close relation between focusing functions and the 
propagator matrix \citep{Wapenaar2022GEO, Wapenaar2023GJI}. Here we briefly review this relation
for a space-dependent material with parameters $\alpha(x)$ and $\beta(x)$. We start by noting that 
the elements $W_x^{P,P}(x,x_0,t)$ and $W_x^{Q,Q}(x,x_0,t)$ are symmetric functions of time, whereas $W_x^{P,Q}(x,x_0,t)$ and $W_x^{Q,P}(x,x_0,t)$ are
asymmetric functions of time. This follows simply from equation (\ref{eq220corJst}) and the expressions for matrices  ${\bf W}_x$ and ${\bf J}$ in equations
(\ref{eq90bc}) and (\ref{eq21a}). For elements $W_x^{P,P}(x,x_0,t)$ and $W_x^{P,Q}(x,x_0,t)$ these symmetry 
properties are also clearly seen in Figures \ref{Fig3}a and \ref{Fig3}b, respectively.
Exploiting these symmetries, Marchenko-type focusing functions can be expressed in terms of the elements of the propagator matrix, according to  \citep{Wapenaar2023GJI}
\begin{eqnarray}
F_x^P(x,x_0,t) &=&W_x^{P,P}(x,x_0,t) - \frac{1}{\eta(x_0)} W_x^{P,Q}(x,x_0,t),\label{eq135}\\
F_x^Q(x,x_0,t) &=&W_x^{Q,P}(x,x_0,t) - \frac{1}{\eta(x_0)} W_x^{Q,Q}(x,x_0,t).\label{eq136}
\end{eqnarray}
From these expressions and equations (\ref{eq114b}) and (\ref{eq90bc}) we obtain the following focusing conditions for $x=x_0$ 
\begin{eqnarray}
F_x^P(x_0,x_0,t) &=&\delta(t),\label{eq135foc}\\
F_x^Q(x_0,x_0,t) &=&- \frac{1}{\eta(x_0)}\delta(t). \label{eq136foc}
\end{eqnarray}
Figure \ref{Fig4} shows an $x,t$-diagram of $F_x^P(x,x_0,t)$ (convolved with a temporal wavelet with a central frequency $\omega_0/2\pi=300$ kHz) 
for a focal point at $x_0=40$ mm. The interpretation is as follows. The four blue arrows in the right-most slab indicate leftward propagating waves
that are emitted into the material from the right, at $x=x_N=200$ mm. After interaction with the boundaries between the homogeneous slabs, 
a single leftward propagating wave arrives at $x=x_0=40$ mm, where it obeys the focusing condition of equation (\ref{eq135foc}). 
The red arrows indicate the rightward propagating scattered part of the focusing function $F_x^P(x,x_0,t)$.

Finally, note that the elements of the propagator matrix can be expressed in terms of the Marchenko-type focusing functions, according to
\begin{eqnarray}
W_x^{P,P}(x,x_0,t) &=&\frac12\{F_x^P(x,x_0,t) + F_x^P(x,x_0,-t)\},\label{eq135q}\\
W_x^{P,Q}(x,x_0,t) &=&-\frac{\eta(x_0)}{2}\{F_x^P(x,x_0,t) - F_x^P(x,x_0,-t)\},\label{eq136q}\\
W_x^{Q,P}(x,x_0,t) &=&\frac12\{F_x^Q(x,x_0,t) - F_x^Q(x,x_0,-t)\},\label{eq137q}\\
W_x^{Q,Q}(x,x_0,t) &=&-\frac{\eta(x_0)}{2}\{F_x^Q(x,x_0,t) + F_x^Q(x,x_0,-t)\}.\label{eq138q}
\end{eqnarray}

\subsection{Time-dependent material}\label{sec12.2}

The relations between the Marchenko-type focusing functions and the propagator matrix for a time-dependent material with parameters $\alpha(t)$ and $\beta(t)$
follow from those in section \ref{sec12.1} by applying the mapping of equation  (\ref{eqmap}), hence
\begin{eqnarray}
F_t^U(x,t,t_0) &=&W_t^{U,U}(x,t,t_0) - \eta(t_0) W_t^{U,V}(x,t,t_0),\label{eq185}\\
F_t^V(x,t,t_0) &=&W_t^{V,U}(x,t,t_0) - \eta(t_0) W_t^{V,V}(x,t,t_0),\label{eq186}
\end{eqnarray}
with focusing conditions for $t=t_0$
\begin{eqnarray}
F_t^U(x,t_0,t_0) &=&\delta(x),\label{eq185b}\\
F_t^V(x,t_0,t_0) &=&- \eta(t_0)\delta(x). \label{eq186b}
\end{eqnarray}
Figure \ref{Fig4} shows an $x,t$-diagram of $F_t^U(x,t,t_0)$ (convolved with a spatial wavelet with a central wavenumber $k_0/2\pi=300*10^3$ km$^{-1}$)
for a focal time at $t_0=40 \,\mu$s. For a correct interpretation, the direction of the blue arrows should be reversed. 
Starting with a leftward propagating wave at $t=t_0=40 \,\mu$s, the response  at $t=t_N=200 \,\mu$s, i.e., $F_t^U(x,t_N,t_0)$,
consists of leftward (blue) and rightward (red) propagating waves. 
Using the mapping of equation (\ref{eqmap}), this response is identical to the focusing function $F_x^P(x_N,x_0,t)$ of the complimentary space-dependent material.
\cite{Manen2024arXiv} exploit this property to design an acoustic space-time material which computes its own inverse.

\section{Conclusions}

We have discussed and compared some fundamental aspects of waves in space-dependent and in time-dependent materials.  
The basic equations for a 1D space-dependent material can be transformed into those for a time-dependent material and vice-versa
by interchanging the space and time coordinates and by applying a specific mapping of wave field components, material parameters and source terms.
When the boundary and initial conditions can be transformed in the same way, then also the solutions of the equations for space-dependent and time-dependent materials
can be transformed into one another by the same mapping. 
When the boundary and initial conditions cannot be transformed in the same way, then the solutions are different.

Green's functions in space-dependent and in time-dependent materials obey the same causality condition (i.e., they are zero before the action of the source) and therefore
they cannot be transformed into one another. 
We have derived a source-receiver reciprocity relation for time-dependent materials, 
which relates a causal Green's function to an acausal Green's function with interchanged time coordinates.
This is different from the source-receiver reciprocity relation for space-dependent materials, 
which interrelates two causal Green's functions with interchanged space coordinates.
We have also derived a new representation for retrieving Green's functions from the space 
correlation of passive measurements in time-dependent materials. Unlike the corresponding 
representation for space-dependent materials it is single-sided, meaning that it suffices to correlate two responses (at two time instants) to sources at a single time instant.

Propagator matrices in space-dependent and in time-dependent materials obey boundary and initial conditions which can be transformed into one another in the same way
as the underlying wave equations. Hence, these propagator matrices are interrelated in the same way. This also applies to representations and reciprocity theorems involving 
propagator matrices, and to Marchenko-type focusing functions, which can be expressed as combinations of elements of the propagator matrix.

\section*{Acknowledgements}

We thank the reviewers and editor for their constructive comments, which helped us to improve this paper. 
Johannes Aichele and Dirk-Jan van Manen acknowledge funding from the Swiss National Science Foundation (SNF, grant 197182).

\appendix

\section{Reflection and transmission coefficients}

\subsection{Space-dependent material}\label{AppA1}

We review reflection and transmission coefficients for a wave incident on a space boundary between two homogeneous time-invariant half-spaces.
Although the derivation is well-known, it also serves as an introduction for the derivation of reflection and transmission coefficients of a time boundary in the next subsection.

For a homogeneous time-invariant source-free material, the wave equation is derived by substituting the constitutive equations (\ref{eq3}) and (\ref{eq4})
with constant parameters $\alpha$ and $\beta$ into equations (\ref{eq1}) and (\ref{eq2}), and subsequently eliminating $Q$ from these equations. We thus obtain
\begin{eqnarray}
\frac{1}{c^2}\partial_t^2P - \partial_x^2P  =0,\label{eq7aww}
\end{eqnarray}
with propagation velocity $c$ given by equation (\ref{eq8a}).
Once $P$ is resolved from equation (\ref{eq7aww}), 
$Q$ follows from $\beta\partial_t Q =- \partial_x P$, and $U$ and $V$ follow from the constitutive equations (\ref{eq3}) and (\ref{eq4}).

\begin{figure}
\centerline{\epsfysize=9. cm \epsfbox{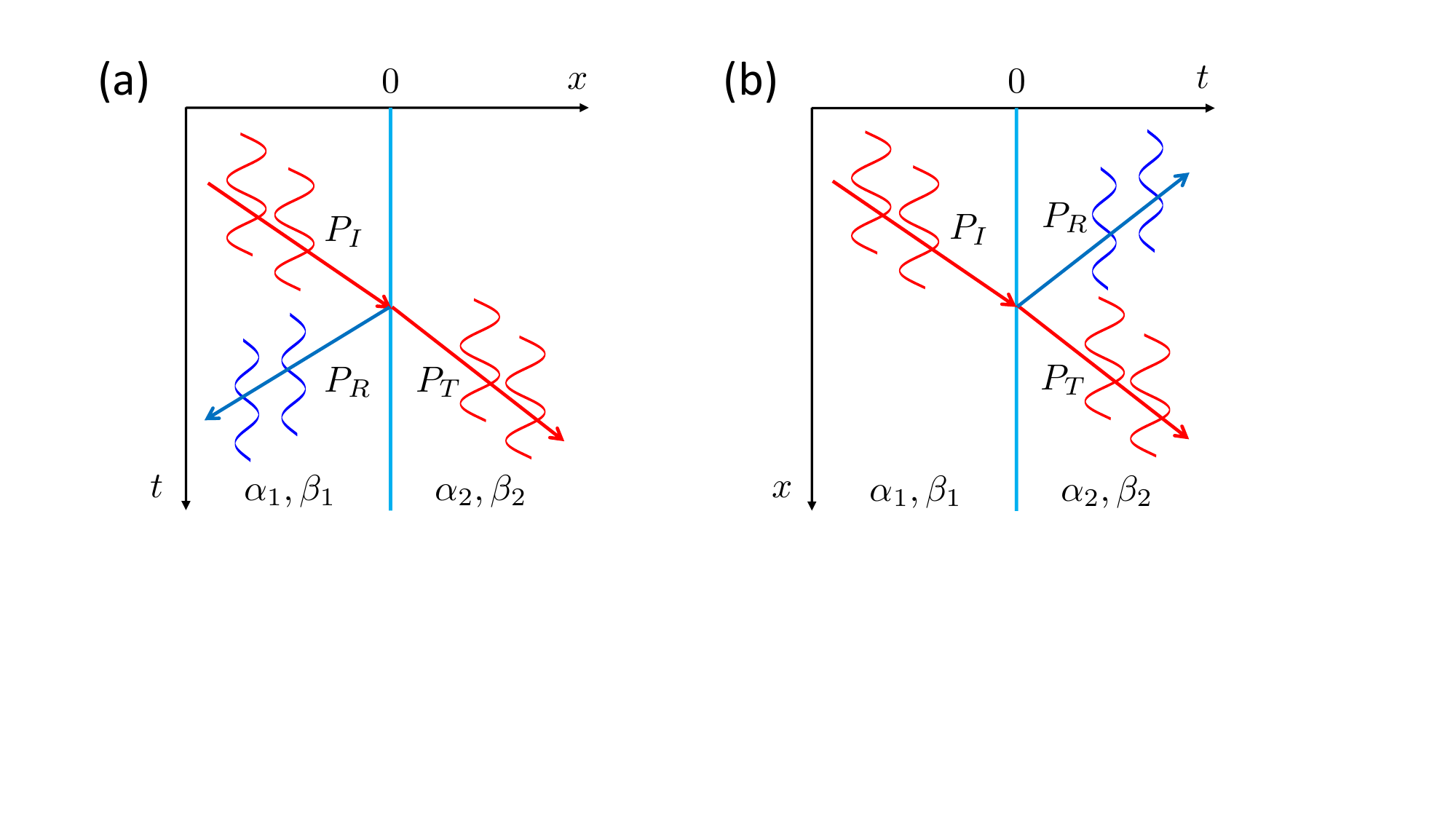}}
\vspace{-3cm}
\caption{Incident ($P_I$), transmitted ($P_T$) and reflected ($P_R$) fields for the situation of a space boundary (a) and a time boundary (b).
Note that the axes are interchanged between (a) and (b). Hence, waves propagating rightward and leftward along the $x$-axis are represented by
rightward and leftward pointing arrows in (a) and by downward and upward pointing arrows in (b).
 }\label{Fig0}
\end{figure}

Consider two homogeneous time-invariant source-free half-spaces, separated by a space boundary, which we take for convenience at $x=0$, see Figure \ref{Fig0}a.
The parameters of the half-spaces $x<0$ and $x>0$ are denoted with subscripts 1 and 2, respectively.
In the half-space $x<0$ we define a rightward propagating monochromatic incident field with unit amplitude and angular frequency $\omega_1$. 
Using complex notation, we have

\begin{eqnarray}
P_I(x,t)&=&\exp i(k_1x-\omega_1t),\quad\mbox{with}\quad  k_1=\frac{\omega_1}{c_1},\label{eq7q}\\
Q_I(x,t)&=&\frac{1}{\eta_1}P_I(x,t),\quad\mbox{with}\quad \eta_1=\beta_1c_1=\sqrt{\frac{\beta_1}{\alpha_1}}.\label{eq8q}
\end{eqnarray}
The rightward propagating transmitted field in the half-space $x>0$ is defined as
\begin{eqnarray}
P_T(x,t)&=&T_x\exp i(k_2x-\omega_2t),\quad\mbox{with}\quad  k_2=\frac{\omega_2}{c_2},\label{eq9q}\\
Q_T(x,t)&=&\frac{1}{\eta_2}P_T(x,t),\quad\mbox{with}\quad  \eta_2=\beta_2c_2=\sqrt{\frac{\beta_2}{\alpha_2}},\label{eq10q}
\end{eqnarray}
where $T_x$ is the transmission coefficient. The subscript $x$ denotes that this coefficient belongs to a space boundary.
The leftward propagating reflected field in the half-space $x<0$ is defined as
\begin{eqnarray}
P_R(x,t)&=&R_x\exp i(-k_1x-\omega_1t),\\
Q_R(x,t)&=&-\frac{1}{\eta_1}P_R(x,t),
\end{eqnarray}
where $R_x$ is the reflection coefficient.
The boundary conditions at $x=0$ state that $P$ and $Q$ are continuous, hence 
\begin{eqnarray}
P_I(0,t)+P_R(0,t)&=&P_T(0,t),\\
Q_I(0,t)+Q_R(0,t)&=&Q_T(0,t).
\end{eqnarray}
These equations should hold for all $t$, from which it follows that $\omega_2=\omega_1$ and hence $k_2=\frac{c_1}{c_2}k_1$,
meaning that the  wavenumber $k_2$ of the transmitted field is different from the wavenumber $k_1$ of the incident and reflected fields 
(unless of course when $c_2=c_1$). 
Moreover, it follows that
\begin{eqnarray}
R_x&=&\frac{\eta_2-\eta_1}{\eta_2+\eta_1},\\
T_x&=&\frac{2\eta_2}{\eta_2+\eta_1},
\end{eqnarray}
which are the well-known expressions for the reflection and transmission coefficients of a space boundary. 
Note that 
\begin{eqnarray}
\frac{1}{\eta_1}(1-R_x^2)=\frac{1}{\eta_2}T_x^2.\label{eq17p}
\end{eqnarray}
We define the net power-flux density in the $x$-direction as 
\begin{eqnarray}
j=\frac12\Re\{P^*Q\}.\label{eq18p}
\end{eqnarray}
Substituting $P=P_I+P_R$, $Q=Q_I+Q_R$ for $x<0$ and $P=P_T$, $Q=Q_T$ for $x>0$ into equation (\ref{eq18p}) we find,  using equation (\ref{eq17p}),
that $j$ is constant. Hence, the net power-flux density of a monochromatic wave field is conserved when traversing a space boundary.

\subsection{Time-dependent material}\label{AppA2}

We review reflection and transmission coefficients for a wave incident on a time boundary between two homogeneous time-invariant ``half-times''
\citep{Xiao2014OL, Morgenthaler58IRE, Mendonca2002PS, Caloz2020IEEE2}.
We take the time boundary for convenience at $t=0$, see Figure \ref{Fig0}b.
The parameters for $t<0$ and $t>0$ are denoted with subscripts 1 and 2, respectively.
The monochromatic incident and transmitted fields are again given by equations (\ref{eq7q}) -- (\ref{eq10q}), this time for $t<0$ and $t>0$, respectively, and with
$T_x$ replaced by $T_t$, with subscript $t$ denoting that this coefficient belongs to a time boundary. Due to causality, the reflected field will not propagate back in time.
Instead, the leftward  propagating reflected field for $t>0$ (indicated by the upward pointing arrow in Figure \ref{Fig0}b) is defined as
\begin{eqnarray}
P_R(x,t)&=&R_t\exp i(k_2x+\omega_2t),\\
Q_R(x,t)&=&-\frac{1}{\eta_2}P_R(x,t),
\end{eqnarray}
where $R_t$ is the reflection coefficient. 
In the literature on waves in time-dependent materials there has been debate whether the tangential electric and magnetic flux densities ($D_y$, $D_z$, $B_y$ and $B_z$)
or the tangential electric and magnetic field strengths ($E_y$, $E_z$, $H_y$ and $H_z$) should be continuous at a time boundary. 
\cite{Xiao2014OL, Morgenthaler58IRE, Mendonca2002PS, Caloz2020IEEE2}  are proponents of continuity of the flux densities and  \cite{Jiang75IEEE} 
is a proponent of continuity of the field strengths. Today the consensus is that the flux densities are continuous at a time boundary, see 
 \cite{Xiao2014OL, Mendonca2002PS, Caloz2020IEEE2} for a clear explanation. 
According to Table \ref{table1} the flux densities are examples of the $U$ and $V$ fields. Taking $U$ and $V$
continuous at $t=0$ we obtain (using equations (\ref{eq3}) and (\ref{eq4}))
\begin{eqnarray}
\alpha_1P_I(x,0)&=&\alpha_2\{P_R(x,0)+P_T(x,0)\},\\
\beta_1Q_I(x,0)&=&\beta_2\{Q_R(x,0)+Q_T(x,0)\}.
\end{eqnarray}
These equations should hold for all $x$, from which it follows that $k_2=k_1$ and hence $\omega_2=\frac{c_2}{c_1}\omega_1$, 
meaning that the  frequency $\omega_2$ of the transmitted and reflected fields is different from the frequency $\omega_1$ of the incident field 
\citep{Apffel2022PRL, Mendonca2002PS, Caloz2020IEEE2} (unless of course when $c_2=c_1$). 
Moreover, it follows that
\begin{eqnarray}
R_t&=&\frac{1}{2}\Bigl(\frac{\alpha_1}{\alpha_2}-\frac{c_2}{c_1}\Bigr),\\
T_t&=&\frac{1}{2}\Bigl(\frac{\alpha_1}{\alpha_2}+\frac{c_2}{c_1}\Bigr).
\end{eqnarray}
These are expressions for the reflection and transmission coefficients of a time boundary.
For TE waves, substituting the material parameters from row 1 of Table \ref{table1}, 
it is easily seen that these expressions are the same as those given by equations (4) and (5) in \cite{Xiao2014OL}.
Note that 
\begin{eqnarray}
\frac{\alpha_1}{c_1}=\frac{\alpha_2}{c_2}(T_t^2-R_t^2).\label{eq24p}
\end{eqnarray}
We define the net field-momentum density in the $x$-direction
as \citep{Burns2020NJP, Feynmann63Book2}
\begin{eqnarray}
M=\frac{1}{2}\Re\{U^*V\}.\label{eq26p}
\end{eqnarray}
Substituting $U=\alpha_1P_I$, $V=\beta_1Q_I$ for $t<0$ and $U=\alpha_2(P_R+P_T)$, $V=\beta_2(Q_R+Q_T)$ 
for $t>0$ into equation (\ref{eq26p}) we find, using equation (\ref{eq24p}), 
that $M$ is constant. Hence, the net field-momentum density of a monochromatic wave field is conserved when traversing a time boundary.

Using equations (\ref{eq3}), (\ref{eq4}) and (\ref{eq8a}), the net power-flux density in the $x$-direction can be expressed as 
$j=c^2M$. Since $M$ is constant, we find $j_2=\frac{c_2^2}{c_1^2}j_1$. 
This is the discrete counterpart of equation (\ref{eq51uj}) for a continuously varying time-dependent material.

\bibliographystyle{gji}

\end{spacing}
\end{document}